\documentclass[
10pt,aps,floatfix,amssymb,prd,twocolumn,
superscriptaddress,nofootinbib,preprintnumbers
]{revtex4-2}

\usepackage{ragged2e}
\usepackage{tabularx}
\usepackage{amssymb,amsmath,verbatim,mathtools}
\usepackage{needspace,enumitem,etoolbox}
\usepackage{graphicx,microtype,afterpage,bm}
\usepackage[dvipsnames]{xcolor}
\usepackage[T1]{fontenc}
\usepackage[utf8]{inputenc}
\usepackage[normalem]{ulem}
\usepackage{booktabs}
\usepackage{multirow}
\usepackage{relsize}
\usepackage{tikz}
\usetikzlibrary{shapes.misc,positioning,arrows.meta}
\usepackage{framed}
\usepackage{nicefrac}
\usepackage{siunitx}

\definecolor{linkcolor}{rgb}{0.0,0.3,0.5}

\usepackage[
unicode,
colorlinks=true,
linkcolor=linkcolor,
citecolor=linkcolor,
filecolor=linkcolor,
urlcolor=blue,
pdfusetitle
]{hyperref}

\usepackage{orcidlink}
\usepackage[capitalise]{cleveref}

\renewcommand\[{\left[}

\def\lsim{\raise 0.4ex\hbox{$<$}\kern -0.8em\lower 0.62
ex\hbox{$\sim$}}

\def\gsim{\raise 0.4ex\hbox{$>$}\kern -0.8em\lower 0.62
ex\hbox{$\sim$}}

\def\lbar{{\hbox{$\lambda$}\kern -0.7em\raise 0.6ex
\hbox{$-$}}}

\newcommand\p{\partial}

\newcommand\ee{\end{equation}}
\newcommand\be{\begin{equation}}
\def\bea{\begin{array}}
\def\eea{\end{array}}\def\ea{\end{array}}
\newcommand\ees{\end{eqnarray}}
\newcommand\bees{\begin{eqnarray}}

\def\dslash{\hspace{-1mm}\not{\hbox{\kern-2pt $\partial$}}}
\def\Dslash{\not{\hbox{\kern-2pt $D$}}}
\def\pslash{\not{\hbox{\kern-2.1pt $p$}}}
\def\kslash{\not{\hbox{\kern-2.3pt $k$}}}
\def\qslash{\not{\hbox{\kern-2.3pt $q$}}}

\def\p1{{\bf p}_1}
\def\p2{{\bf p}_2}
\def\k1{{\bf k}_1}
\def\k2{{\bf k}_2}

\newcommand{\dddM}{\kern 0.2em \raise 1.9ex\hbox{$...$}\kern -1.0em \hbox{$M$}}
\newcommand{\dddQ}{\kern 0.2em \raise 1.9ex\hbox{$...$}\kern -1.0em \hbox{$Q$}}
\newcommand{\dddI}{\kern 0.2em \raise 1.9ex\hbox{$...$}\kern -1.0em\hbox{$I$}}
\newcommand{\dddJ}{\kern 0.2em \raise 1.9ex\hbox{$...$}\kern-1.0em
\hbox{$J$}}
\newcommand{\dddcalJ}{\kern 0.2em \raise 1.9ex\hbox{$...$}\kern-1.0em
\hbox{${\cal J}$}}

\newcommand{\dddO}{\kern 0.2em \raise 1.9ex\hbox{$...$}\kern -1.0em
\hbox{${\cal O}$}}
\def\dddz{\raise 1.5ex\hbox{$...$}\kern -0.8em \hbox{$z$}}
\def\dddd{\raise 1.8ex\hbox{$...$}\kern -0.8em \hbox{$d$}}
\def\dddbd{\raise 1.8ex\hbox{$...$}\kern -0.8em \hbox{${\bf d}$}}
\def\ddbd{\raise 1.8ex\hbox{$..$}\kern -0.8em \hbox{${\bf d}$}}
\def\dddx{\raise 1.6ex\hbox{$...$}\kern -0.8em \hbox{$x$}}

\allowdisplaybreaks

\crefname{section}{Sec.}{Sec.}
\Crefname{section}{Section}{Section}
\crefname{appendix}{App.}{App.}
\Crefname{appendix}{Appendix}{Appendix}
\crefname{footnote}{footnote}{footnote}
\Crefname{footnote}{footnote}{footnote}
\crefname{enumi}{item}{item}
\Crefname{enumi}{Item}{Item}
\crefname{page}{page}{page}
\Crefname{page}{Page}{Page}

\def\pymcpop{\texttt{pymcpop-gw} }

\DeclareSIUnit \parsec {pc}
\DeclareSIUnit \arcsecondfull {arcsec}
\DeclareSIUnit \year {yr}
\DeclareSIUnit \day {day}
\DeclareSIUnit \hour {hr}
\DeclareSIUnit \radiant {rad}
\DeclareSIUnit \degfull {deg}
\DeclareSIUnit \erg {erg}
\DeclareSIUnit \Lsun {L_\odot}
\DeclareSIUnit \Msun {M_\odot}
\DeclareSIUnit \AstroUnit {au}

\newcommand{\dd}[1]{{\rm d}#1}

\newcommand{\mrs}{%
Aix-Marseille Universit\'e, Universit\'e de Toulon,
CNRS, CPT, Marseille, France%
}

\begin{document}

\title{A gaussian--process reconstruction of gravitational--wave friction from dark standard sirens}

 \author{Michele Mancarella\texorpdfstring{$\,$}{ }\orcidlink{0000-0002-0675-508X}}
 \email{michele.MANCARELLA@univ-amu.fr}
\affiliation{\mrs}

\author{Viola De Renzis\texorpdfstring{$\,$}{ }\orcidlink{0000-0001-7038-735X}}
\affiliation{\mrs}

\author{Alessandro Agapito$\,$\orcidlink{0009-0005-9004-3163}
}
\affiliation{\mrs}

\date{\today}

\begin{abstract}
Standard sirens provide a direct probe of the cosmic distance--redshift
relation and therefore a powerful test of whether gravitational waves (GWs)
propagate as predicted by general relativity on cosmological scales.
In many extensions of general relativity, GW propagation is modified by an
anomalous friction term, which alters the amplitude evolution and leads to a
GW luminosity distance that differs from its electromagnetic counterpart.
In this work, we develop a non-parametric reconstruction of this anomalous
friction from dark standard sirens. 
We reconstruct its redshift dependence,
encoded in the friction function $\alpha_M(z)$, with a Gaussian process (GP) and
derive the corresponding GW-to-EM luminosity-distance ratio $\Xi(z)$ by
integration, explicitly retaining the relation between the local friction
modification and its accumulated effect on the observed distance, while
enforcing a monotonic and invertible GW luminosity distance.
We apply the method to the GWTC--5.0 catalog,
jointly inferring the source population, modified propagation, and background
cosmological parameters. We find no evidence for departures from GR, with
$\alpha_M=0$ and $\Xi=1$ consistent with the posterior over the redshift
range probed by the catalog. The result is stable under variations of the GP
kernel, smoothness and amplitude priors, resolution prescription, and
likelihood treatment. We further introduce function-space diagnostics for
comparing flexible and parametric reconstructions, including prior
calibration, prior-to-posterior contraction, common summaries of
propagation--population degeneracies, and principal-component analysis.
These tools make explicit which features are informed by the data and which
are imposed by the function-space prior, providing a framework for
interpretable non-parametric tests of gravitational-wave propagation.
\end{abstract}

\maketitle

\section{Introduction}\label{intro}

Gravitational-wave (GW) observations enable ``standard siren''
cosmology~\cite{Schutz:1986gp,Holz:2005df}: direct measurements of
luminosity distance from compact-binary signals can be combined with redshift
information to infer the expansion history of the Universe. 
A powerful route is provided by spectral sirens, where
features in the source-frame mass distribution of merging black holes break
the intrinsic mass--redshift degeneracy of the GW waveform
\citep{Taylor:2011fs,Farr:2019twy,Mastrogiovanni:2021wsd,
Ezquiaga:2022zkx}. State-of-the-art analyses jointly infer the BBH
population and cosmological parameters, yielding independent measurements of
the Hubble constant~\cite{Mastrogiovanni:2023emh,Gray:2023wgj,
Borghi:2023opd}. 

While much of the standard-siren program has focused on measuring $H_0$,
the same distance--redshift information also provides a direct test of GW
propagation~\cite{Saltas:2014dha,Pettorino:2014bka,Nishizawa:2017nef,Amendola:2017ovw,Belgacem:2017ihm,Belgacem:2018lbp,Lagos:2019kds}. This is especially relevant for the physics of cosmic
acceleration and dark energy, whose origin remains one of the central open
questions in cosmology and is further sharpened by recent large-scale
structure measurements, e.g.~\cite{Ishak:2024jhs}. If gravity departs from
GR on cosmological scales, or if dark energy is dynamical, the amplitude
evolution of GWs can differ from the GR prediction through an additional
friction term in the propagation equation
\cite{Saltas:2014dha,Pettorino:2014bka,Nishizawa:2017nef,
Amendola:2017ovw,Belgacem:2017ihm,Belgacem:2018lbp,Lagos:2019kds}. The distance inferred
from the GW amplitude is then not the standard electromagnetic (EM)
luminosity distance, but a GW luminosity distance
$d_L^{\rm GW}$~\cite{Belgacem:2017ihm,Belgacem:2018lbp,
LISACosmologyWorkingGroup:2019mwx}. Equivalently, modified propagation can
be described by the GW-to-EM distance ratio
$\Xi(z)\equiv d_L^{\rm GW}(z)/d_L^{\rm EM}(z)$, or by the GW extra-friction function
$\alpha_M(z)$. This effect acts on top of possible modifications to the
background expansion and accumulates along the propagation, making it
directly constrainable with standard sirens~\cite{Belgacem:2018lbp}.

Existing constraints on modified GW propagation
\cite{Finke:2021aom,Ezquiaga:2021ayr,Mancarella:2021ecn,Leyde:2022orh,
Mastrogiovanni:2023emh,Chen:2023wpj,LIGOScientific:2025jau} have mostly
adopted theory-motivated parametric forms. One common approach
parameterizes directly the GW-to-EM distance ratio, as in the
$\Xi_0$--$n$ ansatz~\cite{Belgacem:2018lbp}. This form captures broad
classes of luminal modified-gravity models, but does not span the full
theory space, including DHOST, extra-dimensional, and bigravity scenarios
\cite{Langlois:2015cwa,Dvali:2000hr,Hassan:2011zd,
LISACosmologyWorkingGroup:2019mwx}. A second approach parameterizes the
friction function $\alpha_M(z)$ itself, which is directly connected to the
Planck-mass running in Horndeski gravity~\cite{Bellini:2014fua} and appears
more generally as a free function in the effective field theory of dark
energy~\cite{Gleyzes:2014rba}. In practice, one often assumes that
$\alpha_M(z)$ tracks the onset of dark energy, for example by making it
proportional to the background dark-energy density
\cite{Lagos:2019kds,Ezquiaga:2021ayr,Noller:2018wyv,Baker:2020apq,
Mastrogiovanni:2023emh,Chen:2023wpj,Seraille:2024beb,Ishak:2024jhs,
LIGOScientific:2025jau,LIGOScientific:2026uyd}. This choice is well
motivated but not unique, and the impact of such functional assumptions on
the resulting constraints remains debated
\cite{Linder:2015rcz,Linder:2016wqw,Denissenya:2018mqs,Gleyzes:2017kpi}.

This motivates a more general strategy: bypassing a fixed parametric form
and allowing the data to inform the redshift dependence of modified GW
propagation directly. Nonparametric, or weakly modelled, reconstructions
provide a natural framework for this goal; in particular, see Ref.~\cite{Pierra:2025hoc} for a reconstruction of the Hubble function from GW data. 
In the context of modified GW propagation and standard-siren techniques, related approaches have been first suggested in a forecasting context using Gaussian processes with bright sirens~\cite{Belgacem:2019zzu}. Only recently a Fourier-basis reconstruction of the GW-to-EM luminosity-distance ratio with spectral sirens was obtained from LVK catalogs~\cite{Colangeli:2026kbi}.
In this work we suggest a complementary route and introduce a Gaussian-process reconstruction
tailored directly to GW friction. Instead of modelling the GW-to-EM distance ratio as
the primary object, we reconstruct the friction function $\alpha_M(z)$ and
derive the distance ratio from it. This follows the physical structure of
the propagation problem: $\alpha_M(z)$ describes the local modification to
GW propagation, while $\Xi(z)=d_L^{\rm GW}(z)/d_L^{\rm EM}(z)$ is the
cumulative effect measured by standard sirens. The approach therefore keeps
the connection between the propagation equation and the observed
luminosity-distance modification explicit, while still allowing a broad
range of redshift dependences.
After validating the method on simulated cases, we apply it to
the GWTC--5.0 BBH catalog~\cite{LIGOScientific:2026sit} and reconstruct both
$\alpha_M(z)$ and $\Xi(z)$.

A central focus of this work is the interpretation and robustness of the
non-parametric reconstruction. We therefore vary several GP modelling
choices, which act as effective changes of the function-space prior, and test
whether the inferred propagation history is stable. In addition, we introduce
a set of diagnostics to quantify how the data update that prior and to enable
a controlled comparison with parametric models. These include a calibration
of the GP amplitude in function space, prior-to-posterior contraction
measures, common function-space summaries for studying degeneracies, and a
principal-component analysis of the posterior covariance. We use these tools
to compare with the standard $\Xi_0$--$n$ parameterization and to distinguish
information supplied by the catalog from structure imposed by the assumed
functional form.

\section{Gaussian--Process Reconstruction of Modified Gravitational--Wave Propagation}
\label{sec:gp_model}

In several extensions of GR, GWs propagate
with an effective friction term that differs from the one implied by the
standard background expansion.
Here, we consider the propagation equation for the tensor modes
\begin{equation}
h_A''
+
\left[2+\alpha_M(\eta)\right]\mathcal{H}h_A'
+
k^2 h_A
=
0 ,
\label{eq:GW_propagation_general}
\end{equation}
where primes denote derivatives with respect to conformal time $\eta$,
$\mathcal{H}=a'/a$ is the conformal Hubble parameter, $A=+,\times$
labels the two tensor polarizations, and $k$ is the comoving wavenumber.
The function $\alpha_M$ modifies the friction experienced by GWs during
propagation. In Horndeski gravity, with matter minimally coupled to the
Jordan-frame metric, it is the logarithmic running of the effective Planck
mass, $\alpha_M \equiv {\rm d}\ln M_*^2/{\rm d}\ln a$, where $M_*^2$ is the coefficient of the tensor kinetic term
\cite{Bellini:2014fua,Amendola:2017ovw}.
Equation~\eqref{eq:GW_propagation_general} corresponds to the luminal,
massless and source-free limit of a more general tensor propagation
equation; we assumed
$c_T=1$, $m_T=0$, and $\Pi_A=0$, where $c_T$ is the tensor propagation
speed, $m_T$ an effective tensor mass, and $\Pi_A$ a possible source term,
for instance from anisotropic stress
\cite{Ezquiaga:2018btd,LISACosmologyWorkingGroup:2019mwx}.
The assumption $c_T=1$ is motivated by the multimessenger observation of
GW170817, which constrains the relative difference between the GW and
electromagnetic propagation speeds to approximately one part in $10^{15}$~\cite{TheLIGOScientific:2017qsa,LIGOScientific:2017zic}.\footnote{This nearby event does not establish luminality at arbitrary
redshift. However, the strength of the constraint would require a theory to imply a raise in the tensor speed excess of $\sim15$ orders of magnitude in a redshift range between $\sim 0.01$ and $\sim 1$. }
A more fundamental issue is relevant to the luminality assumption. The currently employed spectral-siren likelihood uses the standard mass relation $m_i^{\rm det}=(1+z)m_i^{\rm src}$.
This relation follows from the standard cosmological time dilation $\mathrm{d}t_0=(1+z)\,\mathrm{d}t_s$.
For homogeneous propagation with a time-dependent speed
$c_T(z)$, successive wavecrests instead satisfy
$\mathrm{d}t_0=(1+z)c_T(z)\,\mathrm{d}t_s/c_T(0)$ (see e.g.
Appendix A of Ref.~\cite{LISACosmologyWorkingGroup:2019mwx}), which implies a modified mass scaling. 
Allowing an arbitrary distance-ratio function does not account for this modified
conversion. Consequently, any constraint based on a likelihood making use of the standard source-to-detector frame-mass relation does not cover
time-dependent nonluminal propagation. Such an extension requires a joint
model for mass redshifting and amplitude propagation; we leave this for future work.

The friction term modifies the relation between
the observed GW amplitude and the source redshift, and can be described by
allowing the GW luminosity distance, $d_L^{\rm GW}(z)$, to differ from the
EM luminosity distance, $d_L^{\rm EM}(z)$.
Possible deviations between the GW and EM luminosity distances are modeled
through the distance-ratio function
\begin{equation}
\label{eq:Xi_def}
\Xi(z)
\equiv
\frac{d_L^{\rm GW}(z)}{d_L^{\rm EM}(z)} .
\end{equation}

Under the propagation assumptions adopted above, the anomalous friction is
the only modification of GW propagation, and the GW-to-EM luminosity-distance
ratio is directly related to $\alpha_M(z)$ as
\begin{align}
\frac{{\rm d}}{{\rm d}z}\ln \Xi(z)
&=
\frac{\alpha_M(z)}{2(1+z)} ,
\label{eq:alphaM_Xi_relation}
\\
\Xi(z)
&=
\exp\left[
\frac{1}{2}
\int_0^z
\frac{{\rm d}z'}{1+z'}\,
\alpha_M(z')
\right] .
\label{eq:Xi_alphaM_relation}
\end{align}
Our reconstruction should therefore be understood as a non-parametric
reconstruction within the luminal, massless, friction-modified sector of
GW propagation
\cite{LISACosmologyWorkingGroup:2019mwx}.

For the EM luminosity distance we assume a spatially flat $w_0$CDM background,
\begin{align}
d_L^{\rm EM}(z)
&=
(1+z)d_C(z),
\label{eq:EMdist}
\\
d_C(z)
&=
\frac{c}{H_0}
\int_0^z
\frac{{\rm d}z'}{E(z')},
\end{align}
where
\begin{equation}
E(z)
=
\left[
\Omega_m(1+z)^3
+
(1-\Omega_m)(1+z)^{3(1+w_0)}
\right]^{1/2}.
\end{equation}
Here, $\Omega_{\rm m}$ is the present matter density fraction, and $w_0$ is the dark energy equation-of-state parameter. We also define for later use the logarithmic EM-distance slope
\begin{align}
b(z)
&\equiv
\frac{{\rm d}}{{\rm d}z}
\ln d_L^{\rm EM}(z)
\nonumber\\
&=
\frac{1}{1+z}
+
\frac{1}{d_C(z)}
\frac{{\rm d}d_C}{{\rm d}z}
\nonumber\\
&=
\frac{1}{1+z}
+
\frac{c}{H_0}
\frac{1}{E(z)d_C(z)} .
\label{eq:b_of_z}
\end{align}

A widely used phenomenological parametrization for modified GW propagation is
the two-parameter $\Xi_0$--$n$ model
\cite{Belgacem:2017ihm,Belgacem:2018lbp,LISACosmologyWorkingGroup:2019mwx},
\begin{equation}
\label{eq:Xi0n}
\Xi(z)
=
\Xi_0
+
\frac{1-\Xi_0}{(1+z)^n},
\qquad
\Xi(0)=1,
\qquad
\Xi(z\to\infty)=\Xi_0 .
\end{equation}
This captures a transition from GR at low redshift to an asymptotic constant
ratio at high redshift. Alternatively, one may parametrize $\alpha_M(z)$
directly, for instance by assuming that it scales with the dark-energy density,
$\alpha_M(z)\propto c_M\Omega_{\rm DE}(z)$
\cite{Bellini:2014fua,Lagos:2019kds}. The Gaussian-process reconstruction
introduced below generalizes these parametric families by inferring the
redshift dependence of $\alpha_M(z)$, and therefore of $\Xi(z)$, directly from
the data.

\subsection{Monotonic Gaussian-process construction}
\label{sec:gp_monotonic}

GW observations constrain luminosity distance rather than redshift. The likelihood therefore requires in general the inverse map
$z(d_L^{\rm GW})$, which is well defined only if
$d_L^{\rm GW}(z)$ is strictly increasing over the redshift range of interest.
We enforce this condition by modeling
\begin{equation}
\label{eq:q_of_z}
q(z)
\equiv
\frac{{\rm d}}{{\rm d}z}
\ln d_L^{\rm GW}(z)
\end{equation}
through a positive reparametrization.

We introduce a latent Gaussian process in the coordinate
\begin{equation}
x\equiv \ln(1+z),
\end{equation}
with
\begin{equation}
f(x)
\sim
{\rm GP}\left[0,k(x,x')\right],
\end{equation}
where $k$ is the covariance kernel, and define
\begin{equation}
q(z)
=
\varepsilon_q
+
{\rm softplus}
\left[
\mu_u(z)+f(x)
\right],
\qquad
\varepsilon_q>0 .
\label{eq:q_softplus}
\end{equation}
Here ${\rm softplus}(u)=\ln(1+e^u)$. This mapping guarantees
$q(z)>0$ and hence a monotonic $d_L^{\rm GW}(z)$.

The deterministic centering function $\mu_u(z)$ is chosen so that the prior on $\alpha_M$ and $\Xi$ is
centered on GR. Specifically,
\begin{equation}
\mu_u(z)
=
{\rm softplus}^{-1}
\left[
b(z)-\varepsilon_q
\right],
\label{eq:mu_u_def}
\end{equation}
so that $f\simeq0$ implies
\begin{equation}
q(z)\simeq b(z),
\qquad
\alpha_M(z)\simeq0,
\qquad
\Xi(z)\simeq1,
\end{equation}
and therefore $d_L^{\rm GW}(z)\simeq d_L^{\rm EM}(z)$.

Combining Eqs.~\eqref{eq:Xi_def}, \eqref{eq:alphaM_Xi_relation},
\eqref{eq:b_of_z}, and \eqref{eq:q_of_z}, we obtain
\begin{align}
\alpha_M(z)
&=
2(1+z)
\frac{{\rm d}}{{\rm d}z}
\ln \Xi(z)
\nonumber\\
&=
2(1+z)
\left[
q(z)-b(z)
\right] .
\label{eq:alpha_from_g}
\end{align}
From
Eqs.~\eqref{eq:q_of_z} and \eqref{eq:alpha_from_g}, the monotonicity condition $q(z)>0$
implies
\begin{equation}
\alpha_M(z)
>
-2(1+z)b(z).
\label{eq:alphaM_monotonic_bound}
\end{equation}
This bound is independent of $H_0$, because the explicit dependence on
$H_0$ cancels in the logarithmic derivative defining $b(z)$.

Given $\alpha_M(z)$, or equivalently $q(z)-b(z)$, the distance ratio
$\Xi(z)$ is reconstructed by integrating Eq.~\eqref{eq:Xi_alphaM_relation}.
We impose the local boundary condition $\Xi(0)=1$.

\subsection{Resolution-adapted GP prior}
\label{sec:GP_priors}

Our baseline prior uses a non-stationary Gibbs kernel in
$x=\ln(1+z)$. A stationary kernel assumes a single correlation length across
the full redshift range, so that the prior allows the same characteristic
variation scale at low and high redshift. A Gibbs kernel instead promotes the
correlation length to a function of position, $\ell_{\rm eff}(x)$, and is
therefore suited to a problem in which the resolving power of the data changes
with redshift. This is the case for GW distances: at fixed intrinsic
parameters, the fractional distance uncertainty scales approximately as the
inverse signal-to-noise ratio, and nearby, louder events therefore provide
better distance, and hence redshift, resolution than distant events. We use the Gibbs kernel to encode this information in the prior, allowing shorter correlation lengths where the catalog has better redshift resolution and enforcing smoother behavior where it does not.

Specifically, we write the local correlation length as
\begin{equation}
\ell_{\rm eff}(x)
=
\ell_0\,r(x),
\label{eq:ell_eff_def}
\end{equation}
where $\ell_0$ is a dimensionless smoothness multiplier and $r(x)$ is a
data-driven resolution curve. 

The resolution curve $r(x)$ is constructed from the single-event distance measurement. 
For each event, we draw a fixed number of samples from its luminosity distance posterior, and convert these distances to redshifts
using a fixed reference cosmology with GR propagation. This gives, for event
$i$, a set of reference samples
$\{z_{ij}\}$ and hence
$x_{ij}=\ln(1+z_{ij})$. We assign the event to the median redshift coordinate
\begin{equation}
x_i={\rm quantile}_{0.50}\left(\{x_{ij}\}\right),
\end{equation}
and define its redshift-resolution width as half of the central $68\%$
interval in $x$,
\begin{equation}
w_i
=
\frac{1}{2}
\left[
{\rm quantile}_{0.84}\left(\{x_{ij}\}\right)
-
{\rm quantile}_{0.16}\left(\{x_{ij}\}\right)
\right].
\end{equation}
The pairs $(x_i,w_i)$ are then divided into a fixed number of equally spaced
bins in $x$. In each bin containing a minimum number of events, the bin center
is taken to be the mean of the $x_i$ values in that bin, and the resolution
value $r(x)$ is taken to be a chosen quantile of the corresponding $w_i$
values. In the baseline analysis we use $1000$ reference samples per event,
$15$ requested bins, a minimum of five events per retained bin, and the
$q=0.25$ quantile of the event widths. Thus the baseline resolution curve
tracks the better-measured events in each redshift bin. Variations using a smoothed resolution curve and the
median width, $q=0.50$, are discussed in Sec.~\ref{sec:gwtc5_robustness} below.

The Gibbs covariance is\footnote{In the numerical implementation, we add a small diagonal term,
$10^{-10}\delta_{ij}$, to the GP covariance matrix. This stabilizes the matrix factorization but is small enough not to contribute a physical white-noise component to the GP prior.}
\begin{align}
k_{\rm Gibbs}(x,x')
&=
\eta^2
\left[
\frac{
2\ell_{\rm eff}(x)\ell_{\rm eff}(x')
}{
\ell_{\rm eff}^2(x)+\ell_{\rm eff}^2(x')
}
\right]^{1/2}
\nonumber\\
&\quad\times
\exp\left[
-
\frac{(x-x')^2}
{\ell_{\rm eff}^2(x)+\ell_{\rm eff}^2(x')}
\right] .
\label{eq:gibbs_kernel}
\end{align}

As a robustness test, we also consider a stationary Mat\'ern-$5/2$ kernel,
\begin{align}
k_{\rm Mat}(x,x')
&=
\eta^2
\left(
1+\sqrt{5}\,r
+\frac{5r^2}{3}
\right)
\exp(-\sqrt{5}\,r),
\label{eq:matern52_kernel}
\\
r
&=
\frac{|x-x'|}{\ell}.
\end{align}
The Mat\'ern length-scale prior is chosen to give a smoothness scale comparable
to the baseline Gibbs model in the redshift range most constrained by the data.
This kernel-family variation, together with other changes of the baseline
configuration, is discussed in Sec.~\ref{sec:gwtc5_robustness} below.

\paragraph{Amplitude prior.}
We adopt a penalized-complexity (PC) prior \cite{Simpson:2017pc} for the GP
amplitude $\eta$,
\begin{equation}
p_{\rm PC}(\eta)
=
\lambda_\eta
\exp(-\lambda_\eta\eta),
\qquad
\lambda_\eta
=
\frac{\ln(1/\alpha_\eta)}{U},
\label{eq:eta_pc_prior}
\end{equation}
so that
\begin{equation}
{\rm Pr}(\eta>U)=\alpha_\eta.
\end{equation}
This prior shrinks the reconstruction toward the GR limit, $\eta=0$, while
allowing deviations when supported by the data. The numerical values of
$U$ and $\alpha_\eta$ are specified together with the baseline configuration in
the results section. 

\paragraph{Smoothness prior.}
For the baseline Gibbs model, a smoothness prior is placed on the multiplier
$\ell_0$ in Eq.~\eqref{eq:ell_eff_def}. We use the same PC-prior motivation as
for stationary Mat\'ern fields \cite{Fuglstad:2018pc}: short length scales are
allowed, but penalized unless required by the data. In one dimension this gives
the Fr\'echet form
\begin{equation}
p_{\rm PC}(\ell_0)
=
\frac{1}{2}
\lambda_\ell
\ell_0^{-3/2}
\exp\left[
-\lambda_\ell\ell_0^{-1/2}
\right].
\label{eq:pc_frechet}
\end{equation}
We set $\lambda_\ell$ by imposing
\begin{equation}
{\rm Pr}(\ell_0<\ell_{0,\min})
=
\alpha_{\rm small}.
\label{eq:ell_min}
\end{equation}
The cumulative distribution corresponding to Eq.~\eqref{eq:pc_frechet} is
\begin{equation}
{\rm Pr}(\ell_0<\ell_{0,\min})
=
\exp\left[
-\lambda_\ell\ell_{0,\min}^{-1/2}
\right],
\end{equation}
which gives
\begin{equation}
\lambda_\ell
=
-\ln(\alpha_{\rm small})\,
\ell_{0,\min}^{1/2}.
\label{eq:lambda_ell_def}
\end{equation}

We additionally include a weak exponential penalty against very large values of
$\ell_0$,
\begin{equation}
p(\ell_0)
\propto
p_{\rm PC}(\ell_0)
\exp(-\lambda_{\rm large}\ell_0).
\label{eq:ell_large_penalty}
\end{equation}
This avoids the nearly constant-kernel regime, which can lead to stiff
geometries and inefficient sampling, without imposing a hard upper cutoff. We
parametrize the strength of this penalty through an interpretable upper
smoothness scale,
\begin{equation}
\lambda_{\rm large}
=
\frac{\ln(1/\alpha_{\rm large})}{\ell_{0,\max}} .
\label{eq:lambda_large_def}
\end{equation}
The adopted baseline values and the variations used to test robustness are
given in Sec.~\ref{sec:gwtc5_robustness} below.

\section{Hierarchical Inference and Population Model}
\label{sec:hierarchical_model}

To describe the BBH population underlying the catalog, we adopt a probability
density for source-frame component masses, spins, and redshift,
$p_{\rm pop}(\theta|\Lambda)$, where $\theta$ denotes the physical source
parameters and $\Lambda$ the corresponding population and cosmological
hyperparameters. 

Our primary-mass model follows the default GWTC-4.0 and GWTC-5.0 BBH population model
\cite{LIGOScientific:2025pvj, LIGOScientific:2026ctl}, namely a broken power law plus two Gaussian
components.\footnote{The only intentional modification with respect to the default model of
Ref.~\cite{LIGOScientific:2025pvj} is the functional form of the low-mass
tapering function applied at the primary and secondary low-mass edges, which we
change to improve numerical stability (see~\cite{Agapito:2026kak}). Explicit model definitions are collected
in Appendix~\ref{app:models_nonevol}.} The presence of mass scales in the
population statistically breaks the mass--redshift degeneracy in GW data
and enables cosmological inference. The secondary mass is modeled as a
conditional power law in $m_2$, normalized at fixed $m_1$ and restricted to
$m_2\le m_1$. The mass model is assumed to be redshift-independent (see~\cite{Agapito:2026kak} for a demonstration of the limited impact of redshift evolution). The
merger-rate density is described by a Madau--Dickinson functional form in
redshift. 
In this analysis we do not introduce an explicit spin-population model.\footnote{This means that
the spin distribution is taken to coincide with the spin prior used in the
single-event parameter estimation.}

The full set of parameters and latent variables is denoted by $\Lambda$ and
grouped as
\begin{equation}
\Lambda
=
\left\{
\Lambda_{\rm cosmo},
\Lambda_{\rm MGWP},
\Lambda_z,
\Lambda_m,
\Lambda_\chi
\right\}.
\end{equation}
For the background cosmology we sample
\begin{equation}
\Lambda_{\rm cosmo}
=
\{H_0,\Omega_{\rm m},w_0\}.
\end{equation}
For modified GW propagation, the baseline GP model contains two
hyperparameters $\{\eta,\ell_0\}$, together with a finite set of latent GP coefficients.
In the baseline
configuration we represent the GP on $N_{\rm GP}=212$ nodes in
$x=\ln(1+z)$, sampling the corresponding whitened latent variables $\{f_i\}_{i=1...}^{N_{\rm GP}}$ and
transforming them through the GP covariance.\footnote{We also test a higher-resolution
grid as a numerical robustness check in Sec.~\ref{sec:gwtc5_robustness}.}
Therefore,
\begin{equation}
\Lambda_{\rm MGWP}
=
\{\eta,\ell_0, \{f_i\}_{i=1...}^{N_{\rm GP}}\}.
\end{equation}

For the population, the Madau--Dickinson merger-rate density has three free parameters $\Lambda_z$, while the mass model uses 15 free hyperparameters $\Lambda_m$.

Thus, the model has
\begin{align}
N_{\rm par}
&=
3\;({\rm cosmology})
+
214\;({\rm GP\ hyperparameters})
+
3\;({\rm rate})
\nonumber\\
&\quad
+
15\;({\rm mass}) = 235
\end{align}
degrees of freedom.

The hyperparameters $\Lambda$ are inferred using hierarchical Bayesian inference~\cite{Mandel:2018mve,Vitale:2020aaz}. 
The analysis is based on a rate-marginalized likelihood that incorporates the impact of selection effects. 
Detection efficiency is estimated through reweighted Monte Carlo integration over simulated signals that are recovered by the search pipeline~\cite{Tiwari:2017ndi,Farr:2019rap} with default checks on the Monte Carlo variance of the estimator~\cite{Talbot:2023pex,Heinzel:2025ogf}.

Posterior sampling is performed with an extended implementation of the \texttt{python} package \pymcpop~\cite{pymcpop}, introduced in Ref.~\cite{Mancarella:2025uat}. 
The hierarchical posterior is explored using the No-U-Turn Sampler (NUTS)~\cite{JMLR:v15:hoffman14a,Brooks_2011} within \texttt{numpyro}~\cite{phan2019composable,bingham2019pyro}, with GPU acceleration. 
We monitor the convergence of the chains using the Gelman–Rubin statistic $\hat{r}$~\cite{10.1214/ss/1177011136}, requiring $\hat{r}\approx1$ for all sampled parameters.

A complete description of the likelihood and its numerical implementation is provided in Appendix~\ref{app:stats}. 
Priors are specified in Appendix~\ref{app:priors}.

\section{Validation on simulations}\label{sect:sims}


\begin{figure*}[th]
  \centering

    \includegraphics[width=0.9\textwidth]{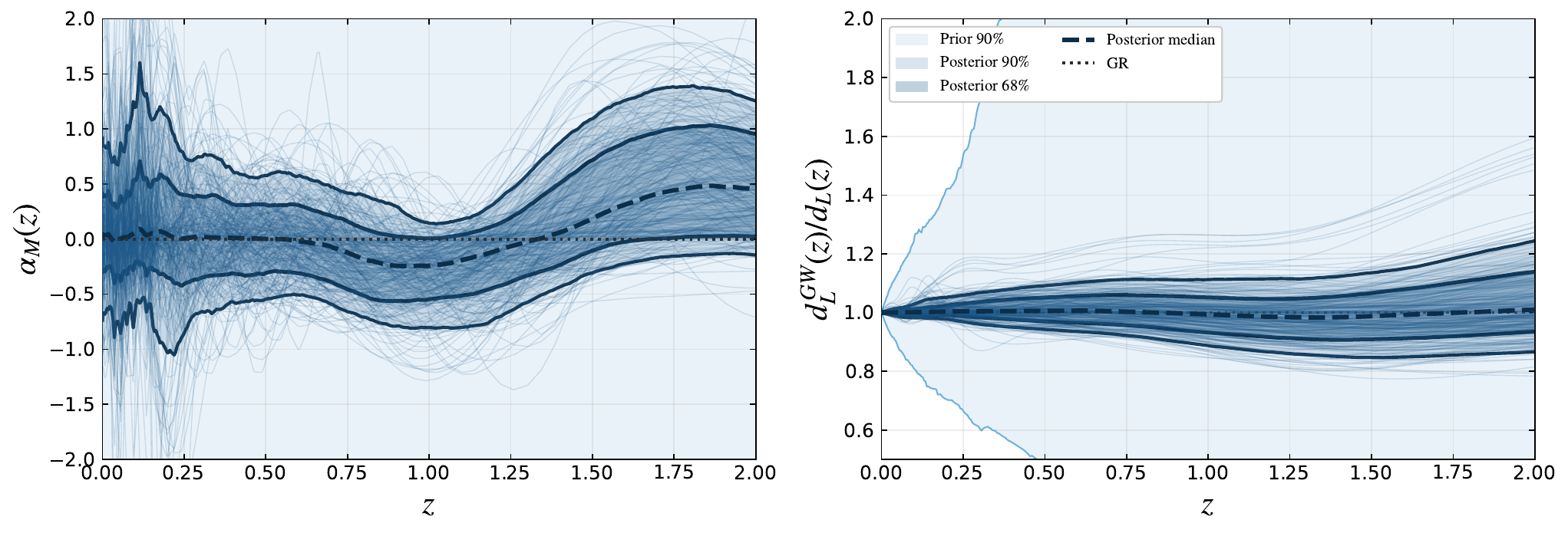}
    
    \includegraphics[width=0.9\textwidth]{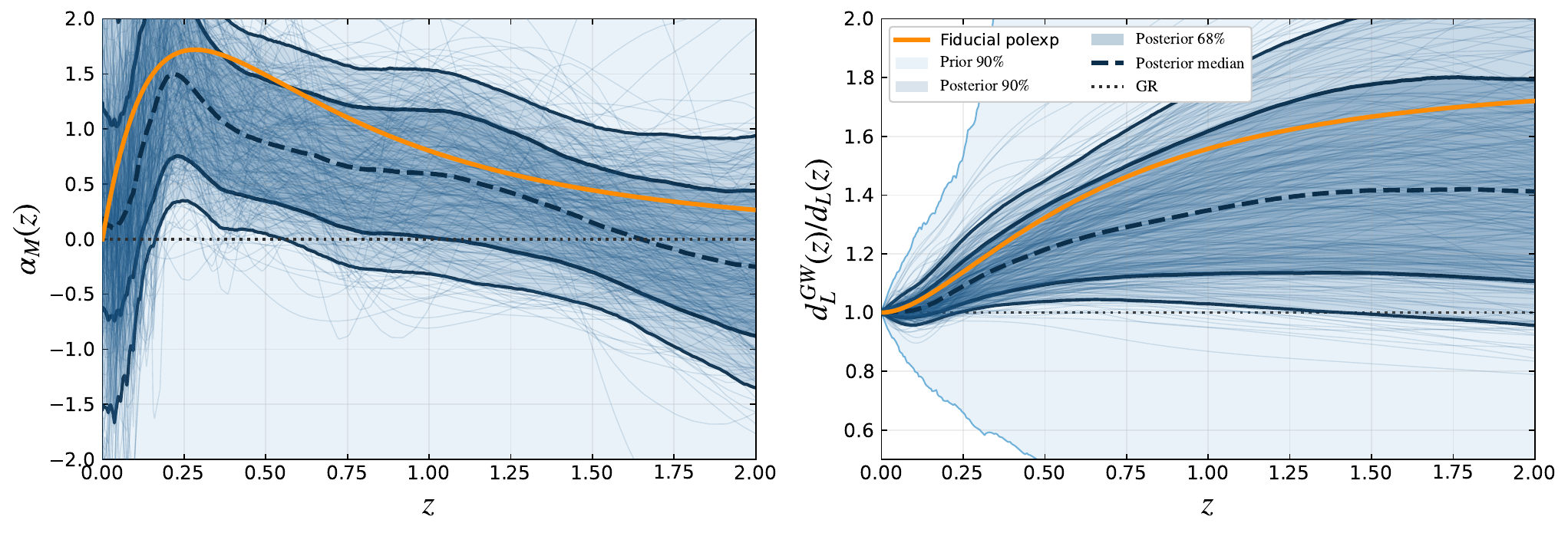}
    
  \caption{Reconstruction of GW friction $\alpha_M(z)$ (left) and the GW-to-EM distance ratio $\Xi(z)$ (right) from 3000 events at O5 design sensitivity. Shaded bands represent prior and posterior credible levels (colors in legend). Top panels: fiducial model is GR (dotted black line). Bottom panels: fiducial model is the DHOST-polexp theory described in the text (solid yellow line).}
  \label{fig:posterior_sims}
\end{figure*}

We first demonstrate that the GP construction is able to pinpoint deviations from GR that are described by more complex functional forms than those covered by standard parameterizations, while also correctly recovering GR in case that is the true underlying model.

We generate simulated catalogs of binary black hole (BBH) mergers assuming the same mass and redshift distributions used in the population model described in App.~\ref{app:models_nonevol}, namely the default "Broken Power-Law plus Two Peaks" of the LVK GWTC-4.0 population paper. Fiducial values correspond to the median posterior of the corresponding data release.
For simplicity we neglect spin degrees of freedom in the simulated sources.

We consider two cosmological scenarios governing gravitational-wave propagation. 
The first corresponds to standard $\Lambda$CDM, in which the GW luminosity distance coincides with the EM distance. 
The second scenario is based on a modified-gravity model that cannot be described by the usual $(\Xi_0,n)$ ansatz or by commonly adopted low-dimensional parameterizations of $\alpha_M$. 
Specifically, we adopt a model representative of DHOST theories~\cite{Langlois:2015cwa}, discussed in Sec.~3.1.2 of Ref.~\cite{LISACosmologyWorkingGroup:2019mwx}, using the polynomial–exponential (``polexp'') parametrization introduced there. 
In this framework modified GW propagation is described through
\begin{equation}
\begin{split}
\Xi(z)=&
\exp\left[
-\frac{(1-\Xi_0)\left[1-(1+z)^n\right]}{(1+z)^{2n}}
\right] \\
\times&\left[\Xi_0+(1-\Xi_0)(1+z)^{-n}\right],
\end{split}
\end{equation}
which leads to the friction term
\begin{equation}
\alpha_M(z)=
-\frac{2n(1-\Xi_0)}{1-\Xi_0+\Xi_0(1+z)^n}
- \frac{2n(1-\Xi_0)}{(1+z)^n}
+\frac{4n(1-\Xi_0)}{(1+z)^{2n}},
\end{equation}
as given in Eqs.~(3.22)–(3.23) of Ref.~\cite{LISACosmologyWorkingGroup:2019mwx}. 
For illustration we choose parameters corresponding to a large deviation from general relativity, $\Xi_0=1.8$ and $n=3$. While not completely ruled out by current observations, these are already in mild tension with current constraints~\cite{LIGOScientific:2026uyd}.
This intentionally extreme choice provides a useful stress test of the reconstruction method, demonstrating that the Gaussian-process model can depart significantly from the GR-centered prior when supported by the data.

Mock observations and injection sets are generated assuming the LVK O5 design sensitivity. 
We use a modified version of the \texttt{GWMockCat} package~\cite{Farah:2023vsc}.
We analyze 3000 events, in order to make the effect of modified gravity distinguishable and to stress test the recovery of GR in a regime where the statistical uncertainty might not be large enough to mask possible residual biases induced by the GP construction.

Figure~\ref{fig:posterior_sims} shows the reconstruction of
$\alpha_M(z)$ (left) and $\Xi(z)$ (right) for the fiducial
$\Lambda$CDM scenario (top) and for the polexp modified-propagation
scenario (bottom). For these tests we adopt the same GP amplitude setting,
$U=8$, used in the GWTC--5.0 analysis described in the next section.
Dark shaded bands show the $50\%$ and $90\%$ posterior credible intervals,
while the lighter region shows the $90\%$ prior range. In the modified
scenario, the injected polexp functions are shown in dark yellow.

For the GR injection, the reconstruction remains centered around
$\alpha_M=0$ and $\Xi=1$, with the posterior contracting relative to the
prior over the redshift range where the simulated catalog is informative.
For the polexp injection, the GP is able to depart substantially from its
GR-centered prior and follows the main redshift dependence of both the local
friction function and the accumulated distance ratio. As expected,
$\alpha_M(z)$ is reconstructed less tightly and exhibits more local
freedom, whereas the integrated quantity $\Xi(z)$ is smoother and more
directly constrained.

These simulations are intended as validation tests of the reconstruction
rather than forecasts for a particular observing run or estimates of the
detectability of modified gravity. The catalog size, detector sensitivity,
and injected departure from GR were chosen to provide informative tests of
whether the method can recover both the GR limit and a non-standard
propagation history outside the usual $\Xi_0$--$n$ functional form. A
dedicated forecast of detection sensitivity is beyond the scope of this work.

\section{Results with GWTC-5.0}\label{sect:gwtc4}

\begin{figure*}[t]
  \centering
  \includegraphics[width=\textwidth]{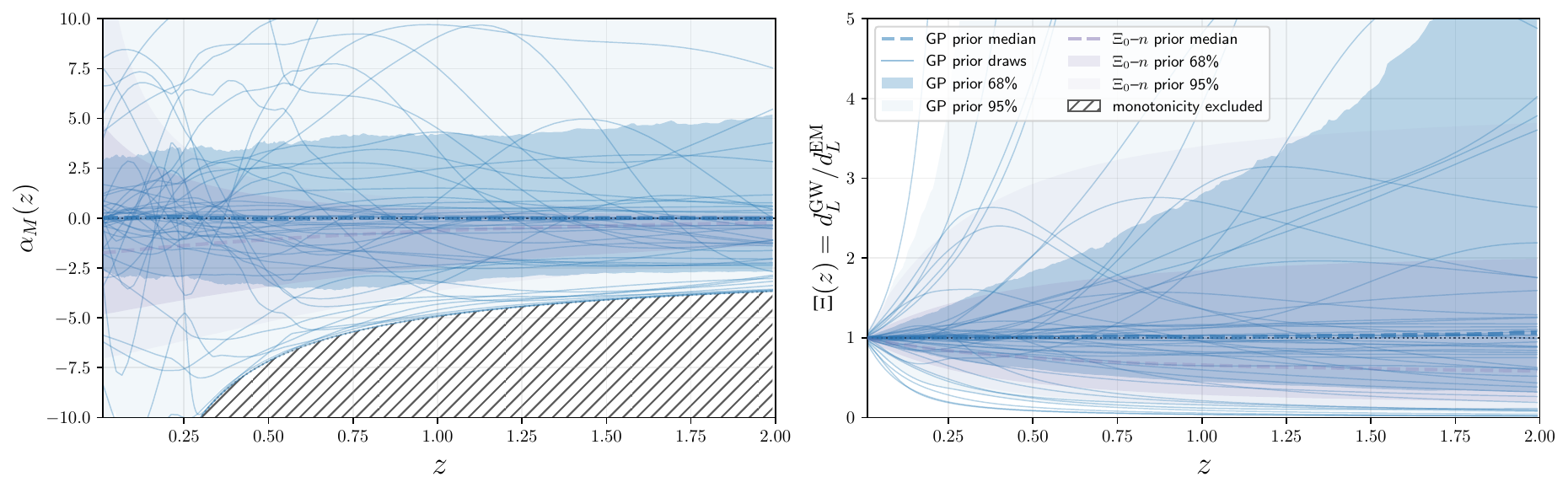}
  \caption{Prior predictive distribution for the GWTC--5 analysis for the GW friction
$\alpha_M(z)$ (left) and the GW-to-EM luminosity-distance
ratio $\Xi(z)$ (right). Blue bands and thin blue curves
show the GP prior predictive distribution and individual GP prior realizations for the selected amplitude hyperparameter $U=8$. Purple bands and dashed purple curves show the induced prior of the standard
$\Xi_0$--$n$ parameterization. 
The hatched region in the $\alpha_M$ panel is
excluded by the monotonicity condition.
}
  \label{fig:prior}
\end{figure*}

We now turn to applying the GP reconstruction  to data from the GWTC-5.0 catalog~\cite{LIGOScientific:2026wfs,LIGOScientific:2026sit}.

\subsection{Data}\label{sec:data_nonevol}

We analyze the GWTC-5.0 catalog following the event selection of the GWTC-5.0 population analysis~\cite{LIGOScientific:2026ctl}.
Specifically, we use BBH events satisfying ${\rm FAR}<1\,{\rm yr}^{-1}$ in the LVK search pipelines.
We use samples obtained with the \texttt{IMRPhenomXPHM} waveform model for events in O1 to O3 runs, and \texttt{IMRPhenomXPHM\_SpinTaylor} for O4a-O4b events.
The data products are available at~\cite{ligo_scientific_collaboration_and_virgo_2025_17014085, ligo_scientific_collaboration_and_virgo_2023_8177023, ligo_scientific_collaboration_and_virgo_2022_6513631,ligo_scientific_collaboration_and_virgo_2026_20348005}.
Selection effects are computed from the GWTC-5.0 injection campaigns, available at~\cite{ligo_scientific_collaboration_2026_19500052} (see ~\cite{Essick:2025zed} for a more complete description and~\cite{Essick:2023toz,PhysRevD.108.043011} for additional details).

\begin{figure*}[t]
  \centering
  \includegraphics[width=\textwidth]{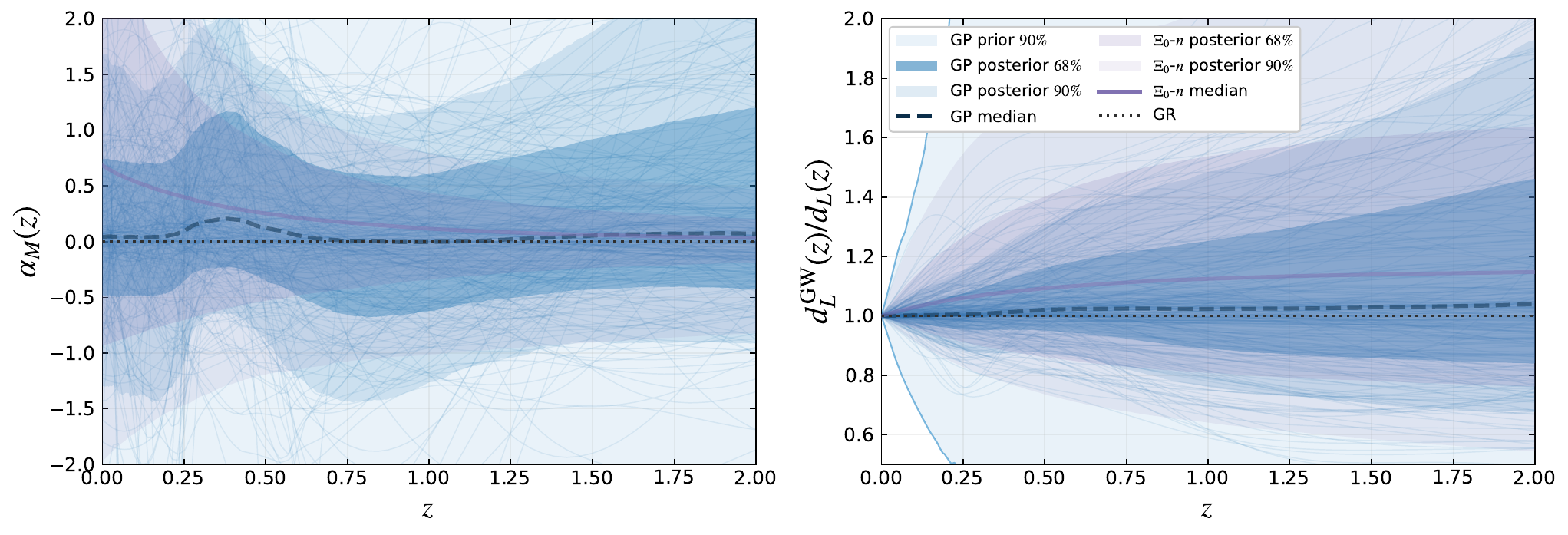}
  \caption{
  Posterior reconstruction of modified GW propagation from the GWTC--5
  catalog, compared with the standard $\Xi_0$--$n$ parameterization. The
  left panel shows the GW friction function $\alpha_M(z)$, and the right
  panel shows the GW-to-EM luminosity-distance ratio
  $d_L^{\rm GW}(z)/d_L(z)\equiv\Xi(z)$. Blue bands show the calibrated GP
  posterior credible regions, while the pale blue band shows the GP prior
  predictive $90\%$ region. Thin blue curves are GP posterior draws. Purple
  bands and the solid purple curve show the posterior induced by the
  $\Xi_0$--$n$ parameterization. The dashed dark-blue curve marks the GP
  posterior median, and the dotted black line denotes the GR prediction,
  $\alpha_M=0$ and $\Xi=1$.
  }
  \label{fig:posterior_GWTC5}
\end{figure*}

\subsection{Prior calibration}
\label{sec:gwtc5_prior_calibration}

We calibrate the GP amplitude prior
to the redshift range effectively probed by this data set. The GP prior uses the resolution curve inferred from the catalog, and we choose the amplitude hyperparameter $U$ by matching the integrated function-space prior width to that induced by the standard $\Xi_0$--$n$ parameterization. 

The priors adopted in this work have support over $0.05<\Xi_0<5$ and $0.05<n<5$, with $\Xi_0$ sampled through a bounded transformation in logarithmic space. The specific prior distributions and their implementation are given in Appendix~\ref{app:priors}.\footnote{Note that this prior is not perfectly GR-centered as is the case of the GP; however, we find the effect to be negligible on the posterior.}

We calibrate the GP amplitude to obtain a comparable integrated prior width in event-supported $\log\Xi$ space. This calibration matches the overall amplitude of the functional uncertainty, but does not impose identical prior distributions or correlations across redshift.

We define equal-populated event-redshift bins $\{B_b\}$ from the event-redshift posteriors, obtained with a reference Planck cosmology.
For each prior realization we compute
\begin{equation}\label{eq:Yb}
Y_b \equiv
\frac{1}{\Delta z_b}
\int_{B_b} dz\,\log\Xi(z)\,
\end{equation}
over the redshift interval $\Delta z_b$,
and summarize the prior width in each bin by
\begin{equation}\label{eq:W68}
W_{68}(Y_b) \equiv Q_{84}(Y_b)-Q_{16}(Y_b).
\end{equation}
We then define the integrated prior-width metric
\begin{equation}
A_{68}\equiv
\left[
\frac{1}{N_{\rm bin}}
\sum_b W_{68}^2(Y_b)
\right]^{1/2}.
\label{eq:A68_prior_calibration}
\end{equation}
Scanning the GP amplitude hyperparameter over
\begin{equation}
U=\{2,4,6,8,10,12\},
\end{equation}
we select the largest scanned value of $U$ for which
$A_{68}^{\rm GP}(U)$ does not exceed the corresponding
$A_{68}^{\Xi_0-n}$ value. This gives $U=8$ for the baseline analysis.

All other GP hyperparameters are kept fixed to the baseline Gibbs-resolution
configuration:
$ \alpha_\eta=0.01,\qquad
\ell_{0,\min}=1.5,\qquad
\ell_{0,\max}=3.5,\qquad
\alpha_{\rm small}=0.01,\qquad
\alpha_{\rm large}=0.05$ .
Figure~\ref{fig:prior} shows the resulting catalog-calibrated GP prior together with the induced $\Xi_0$--$n$ prior.

The hatched region in the $\alpha_M$ panel of Fig.~\ref{fig:prior} shows the
region excluded by monotonicity of the GW luminosity distance, as quantified by Eq.~\eqref{eq:alphaM_monotonic_bound}.


\subsection{Results}
\label{sec:gwtc5_results}

Figure~\ref{fig:posterior_GWTC5} shows the reconstruction of modified GW
propagation from the GWTC--5 catalog. The left panel shows the
friction function $\alpha_M(z)$, while the right panel shows the accumulated
distance ratio $\Xi(z)=d_L^{\rm GW}(z)/d_L(z)$. The GP posterior is
consistent with GR over the full event-supported redshift range: both
$\alpha_M=0$ and $\Xi=1$ lie well within the posterior credible regions.
The posterior is nevertheless visibly contracted relative to the prior,
especially for $\Xi(z)$, which is the quantity directly constrained by the
luminosity-distance information. 

The two panels also illustrate the difference between the local and
accumulated descriptions of modified propagation. The friction function
$\alpha_M(z)$ is a local quantity and therefore allows larger pointwise
fluctuations. By contrast, $\Xi(z)$ is obtained by integrating the propagation
modification and is correspondingly smoother. 

The same figure compares the GP reconstruction with the posterior obtained
from the standard $\Xi_0$--$n$ parameterization. The two reconstructions are
mutually consistent and both are compatible with GR. However, they distribute
the allowed departures from GR differently. The $\Xi_0$--$n$ model induces a
smooth, coherent trend in $\Xi(z)$ and correspondingly gives a positive
$\alpha_M(z)$ already at low redshift. This behaviour is a consequence of the
parameterization: $\alpha_M(z)$ is not independently reconstructed in the
$\Xi_0$--$n$ model, but is derived from the assumed global functional form of
$\Xi(z)$. The GP reconstruction, instead, allows more local variation
through the prior on the latent field governing $q(z)$, constrained through the integrated distance relation. This allows the friction
function to remain closer to GR near $z\simeq0$, while still permitting large
accumulated deviations in the distance ratio.\\

\paragraph{Integrated width and prior-to-posterior contraction.}

To quantify the information gained by the data, we use the same
event-binned $\log\Xi$ metric introduced for the prior calibration. For
each draw we use the quantities defined in Eqs.~\ref{eq:Yb}-\ref{eq:W68}-\ref{eq:A68_prior_calibration}. In particular, $A_{68}$ measures the typical prior or posterior uncertainty in the
event-supported part of $\log\Xi(z)$.

\begin{table}[t]
\centering
\begin{tabular}{lccc}
\hline
Model
& $A_{68}^{\rm prior}$
& $A_{68}^{\rm post}$
& $A_{68}^{\rm prior}/A_{68}^{\rm post}$ \\
\hline
GP
& 1.110
& 0.213
& 5.21 \\
$\Xi_0$--$n$
& 1.264
& 0.448
& 2.82 \\
\hline
\end{tabular}
\caption{
Integrated prior and posterior widths in event-binned $\log\Xi$ space.
The GP amplitude prior was calibrated so that the GP and $\Xi_0$--$n$
priors have comparable integrated widths in this metric. After conditioning
on the GWTC--5 data, the GP posterior contracts more strongly and has an
integrated posterior width about a factor of two smaller than the
$\Xi_0$--$n$ posterior.
}
\label{tab:A68_widths_GWTC5}
\end{table}

The result is reported in Table~\ref{tab:A68_widths_GWTC5}. The two priors
are comparable in the calibration metric, with
$A_{68}^{\rm prior}=1.110$ for the GP and $1.264$ for
$\Xi_0$--$n$. After conditioning on the data, however, the posterior
widths differ more substantially: $A_{68}^{\rm post}=0.213$ for the GP
and $0.448$ for $\Xi_0$--$n$. Thus, after prior calibration, the GP
contracts by a factor $5.21$, compared to $2.82$ for the
$\Xi_0$--$n$ model. This shows that the smaller GP posterior width is
not simply the result of assigning it a narrower prior amplitude. Instead,
the two models organize the information in the catalog differently.\\

\paragraph{Propagation--rate degeneracy.}

A useful way to understand this difference is to inspect how modified
propagation correlates with the population model. In the $\Xi_0$--$n$
parameterization, one can directly look at correlations involving the
parameter $\Xi_0$. For the GP there is no analogous single propagation
parameter. We therefore compress each reconstructed function into a common
function-space summary,
\begin{equation}
\Delta\langle\log\Xi\rangle_{\rm high-low}
\equiv
\langle\log\Xi\rangle_{\rm high}
-
\langle\log\Xi\rangle_{\rm low},
\end{equation}
where the two averages are taken over the highest and lowest event-supported
redshift bins. This quantity measures the net redshift tilt of the accumulated
propagation modification across the range where the catalog carries
information.
In the limit $z_{\rm low}\to0$, $z_{\rm high}\to\infty$, the summary
reduces in the $\Xi_0$--$n$ parameterization to $\log\Xi_0$, independently of $n$.

\begin{figure}[t]
  \centering
  \includegraphics[width=0.9\columnwidth]{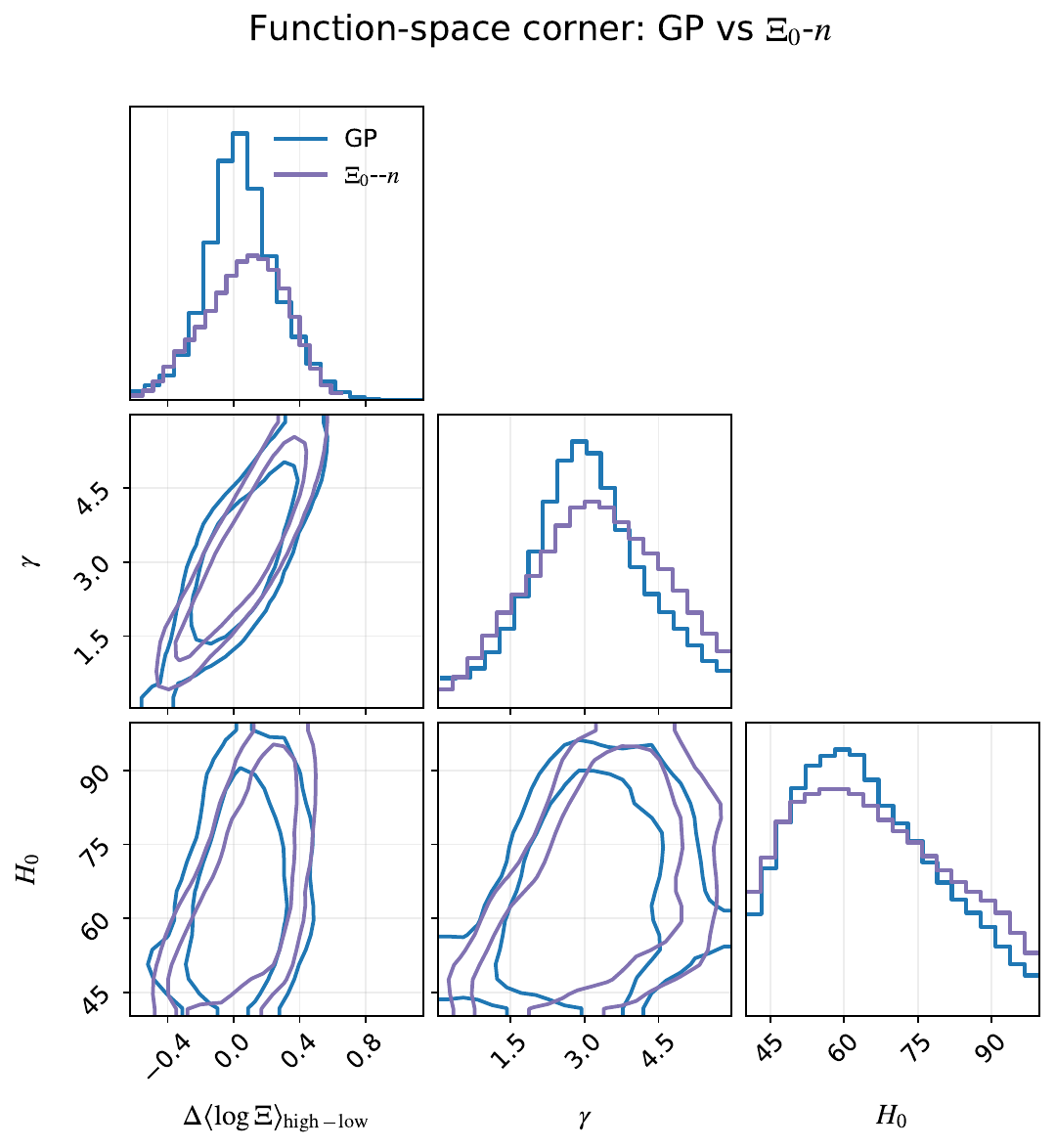}
  \caption{
  Function-space posterior corner for the GP and $\Xi_0$--$n$
  reconstructions. The propagation variable is
  $\Delta\langle\log\Xi\rangle_{\rm high-low}$, the difference between the
  high- and low-redshift averages of $\log\Xi(z)$ over the event-supported
  range. We compare this common propagation summary with the
  Madau--Dickinson rate parameter $\gamma$ and with $H_0$. Both models show
  the same qualitative propagation--rate degeneracy direction.
  }
  \label{fig:function_space_corner_GWTC5}
\end{figure}

Figure~\ref{fig:function_space_corner_GWTC5} shows the joint posterior of
this summary with the Madau--Dickinson rate parameter $\gamma$ and with
$H_0$. Both reconstructions show a strong propagation--rate degeneracy:
a change in the redshift evolution of the source population can be partially
compensated by a change in the redshift dependence of the GW distance ratio.
This degeneracy is therefore not removed by the GP. Rather, the GP expresses
it in a function-space variable instead of a single parametric amplitude.
Numerically, the correlations with the propagation tilt are comparable in
the two models,
\begin{equation}
{\rm corr}\!\left(
\Delta\langle\log\Xi\rangle_{\rm high-low},\gamma
\right)
\simeq
0.84\quad{\rm (GP)},\quad
0.90\quad(\Xi_0\text{--}n),
\end{equation}
while the correlation with $H_0$ is weaker for the GP,
\begin{equation}
{\rm corr}\!\left(
\Delta\langle\log\Xi\rangle_{\rm high-low},H_0
\right)
\simeq
0.28\quad{\rm (GP)},\quad
0.54\quad(\Xi_0\text{--}n).
\end{equation}
The important point is therefore not that the GP eliminates the degeneracy,
but that the degeneracy does not force the GP posterior into the same global
shape as the $\Xi_0$--$n$ parameterization.\\

\paragraph{Function-space covariance structure.}

To
understand more generally how posterior uncertainty is distributed across
redshift, we perform a principal-component analysis (PCA) of the covariance matrix
of the event-binned variables $Y_b=\langle\log\Xi\rangle_b$. This asks
whether the posterior uncertainty is effectively controlled by a single
coherent mode, or whether it contains additional independent directions in
function space.

\begin{table}[t]
\centering
\begin{tabular}{lccccc}
\hline
Model
& PC1
& PC1+PC2
& $r_{\rm eff}$
& ${\rm corr}({\rm PC1},\gamma)$ \\
\hline
GP
& 0.883
& 0.972
& 1.56
& 0.761 \\
$\Xi_0$--$n$
& 0.992
& 1.000
& 1.05
& 0.885 \\
\hline
\end{tabular}
\caption{
PCA summary of the posterior covariance of the event-binned
$\log\Xi$ variables $Y_b$. PC1 and PC1+PC2 give the fraction of total
posterior variance explained by the first one and two principal components.
The effective rank $r_{\rm eff}$ is computed from the entropy of the
eigenvalue spectrum.  
The $\Xi_0$--$n$ posterior is nearly rank-one
in this function space, while the GP retains a higher-rank covariance
structure.
}
\label{tab:pca_function_space_GWTC5}
\end{table}

The PCA results are summarized in Table~\ref{tab:pca_function_space_GWTC5}.
Let $\lambda_i$ denote the eigenvalues of the posterior covariance matrix,
ordered from largest to smallest. The entries labelled PC1 and PC1+PC2 give
the fractions of the total posterior variance carried by the first principal
component and by the first two components, respectively,
\begin{equation}
f_1=\frac{\lambda_1}{\sum_i\lambda_i},
\qquad
f_{1+2}=\frac{\lambda_1+\lambda_2}{\sum_i\lambda_i}.
\end{equation}
Thus, values close to unity indicate that the posterior uncertainty is
concentrated in only one or two coherent redshift-dependent modes.

To summarize the full eigenvalue spectrum with a single number, we also
report the effective rank
\begin{equation}
r_{\rm eff}
=
\exp\left[
-\sum_i p_i\ln p_i
\right],
\qquad
p_i\equiv\frac{\lambda_i}{\sum_j\lambda_j}.
\end{equation}
This quantity can be interpreted as the effective number of principal
components that contribute appreciably to the posterior covariance:
$r_{\rm eff}=1$ corresponds to an exactly rank-one covariance, while
$r_{\rm eff}=2$ would correspond, for example, to two equally important
modes. It need not be an integer.
Finally, ${\rm corr}({\rm PC1},\gamma)$ is the posterior correlation between
the amplitude of the leading principal component and the
Madau--Dickinson rate parameter $\gamma$. It therefore measures how strongly
the dominant propagation mode is coupled to the inferred redshift evolution
of the merger rate.

For the $\Xi_0$--$n$ model, the first principal component accounts for
$99.2\%$ of the posterior variance, and the effective rank is
$r_{\rm eff}=1.05$. This means that the posterior uncertainty is almost
entirely a single coherent deformation of $\log\Xi(z)$ across all redshift
bins. This is the expected behaviour for a fixed two-parameter functional
form.

The GP posterior is also dominated by a leading mode, but less extremely:
the first component accounts for $88.3\%$ of the variance, the first two
components for $97.2\%$, and the effective rank is
$r_{\rm eff}=1.56$. The GP therefore retains additional function-space
directions beyond a single global tilt. This provides a quantitative
explanation for the stronger contraction reported in
Table~\ref{tab:A68_widths_GWTC5}: after calibrating the prior amplitude, the
GP can allocate posterior freedom across redshift more flexibly, whereas the
$\Xi_0$--$n$ model ties the accumulated distance modification to a
nearly rank-one global shape.

Overall, the calibrated GP and the $\Xi_0$--$n$ parameterization give
consistent reconstructions and both are compatible with GR. The difference
is in how the available information is organized. The $\Xi_0$--$n$
model provides a compact global description of the accumulated distance
ratio. The GP reconstruction keeps the physical relation between local
friction and accumulated distance explicit and allows the posterior
uncertainty to occupy a richer, though still data-constrained,
function space.


\begin{table*}[t]
\centering
\small
\begin{tabular}{lll}
\hline
Variant & Change & Test \\
\hline
Baseline & Calibrated Gibbs GP & Reference run \\
Wider $\ell_0$ & Broader smoothness prior & Smoothness sensitivity \\
Smoothed res. & Smoothed resolution curve & Resolution modelling \\
Median res. & Median resolution estimate & Resolution quantile \\
Mat\'ern-5/2 & Stationary Mat\'ern kernel & Kernel choice \\
$U=12$ & Broader amplitude prior & Prior amplitude \\
Lower $\ell_0$ & Shorter correlation scale & Local redshift structure \\
Marginal likelihood & Alternative likelihood & Likelihood implementation \\
\hline
\end{tabular}
\caption{
Robustness variations used for the GWTC--5 GP reconstruction. Each run
changes one modelling choice relative to the calibrated baseline.
}
\label{tab:gwtc5_robustness_variants}
\end{table*}

\subsection{Robustness to GP modelling choices}
\label{sec:gwtc5_robustness}

We test the stability of the GWTC--5 reconstruction by repeating the
analysis under several controlled changes of the GP model, which probe different aspects of the effective function-space prior.
Table~\ref{tab:gwtc5_robustness_variants} summarizes the robustness runs.
Each variant changes one aspect of the GP reconstruction relative to the
baseline, and is intended to test whether the inferred propagation history is
stable under the corresponding modelling choice.

\begin{figure*}[t]
  \centering
  \includegraphics[width=\textwidth]{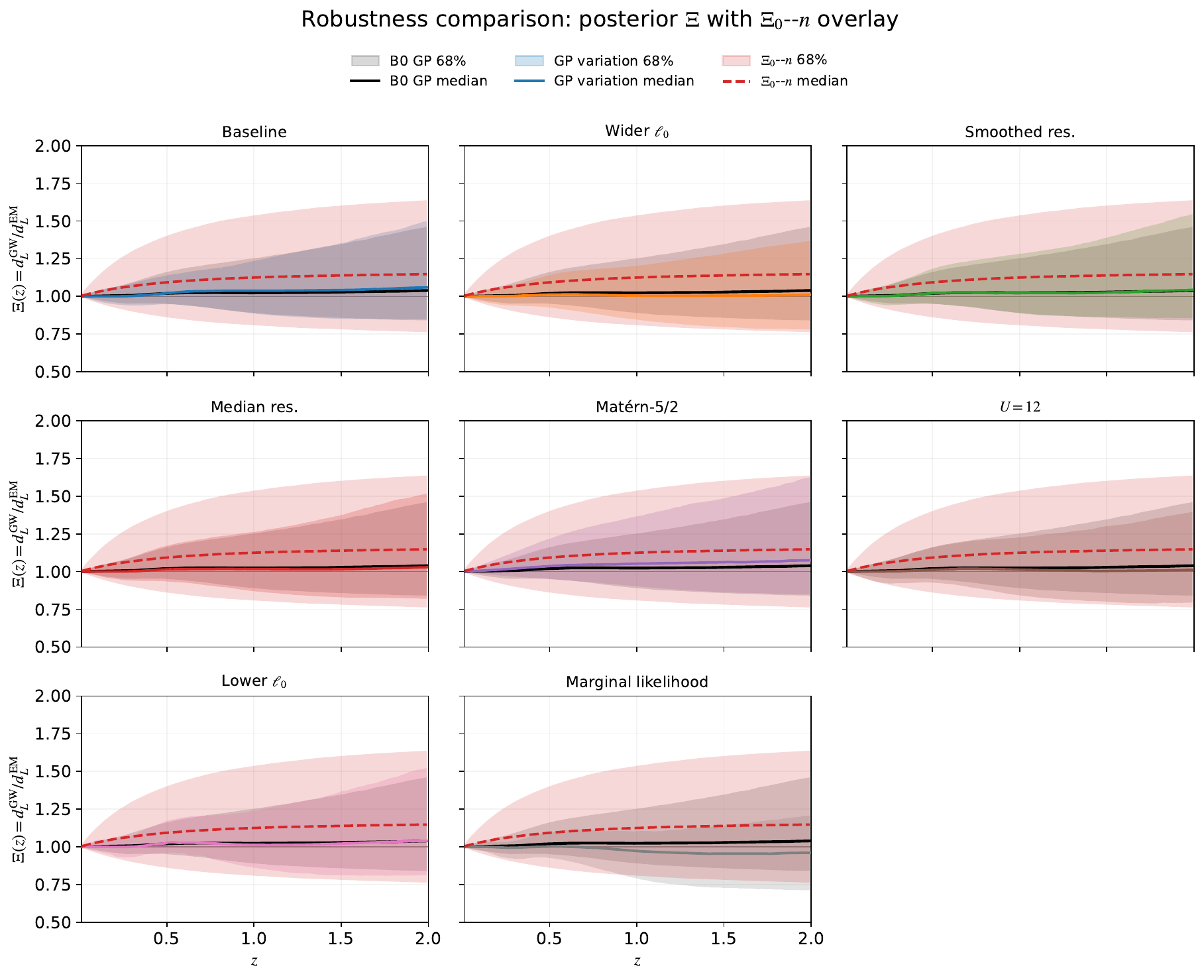}
  \caption{
  Robustness of the reconstructed distance ratio $\Xi(z)$ under GP modelling
  variations. Each panel shows the calibrated baseline GP posterior in gray,
  one GP variant in color, and the $\Xi_0$--$n$ posterior in red. Bands show
  the $68\%$ credible regions and curves show posterior medians. The
  $\Xi_0$--$n$ curve is repeated in each panel as a fixed parametric
  reference. The GP reconstruction of $\Xi(z)$ is stable across the
  variations and remains consistent with GR.
  }
  \label{fig:robustness_Xi_GWTC5}
\end{figure*}

\begin{figure*}[t]
  \centering
  \includegraphics[width=\textwidth]{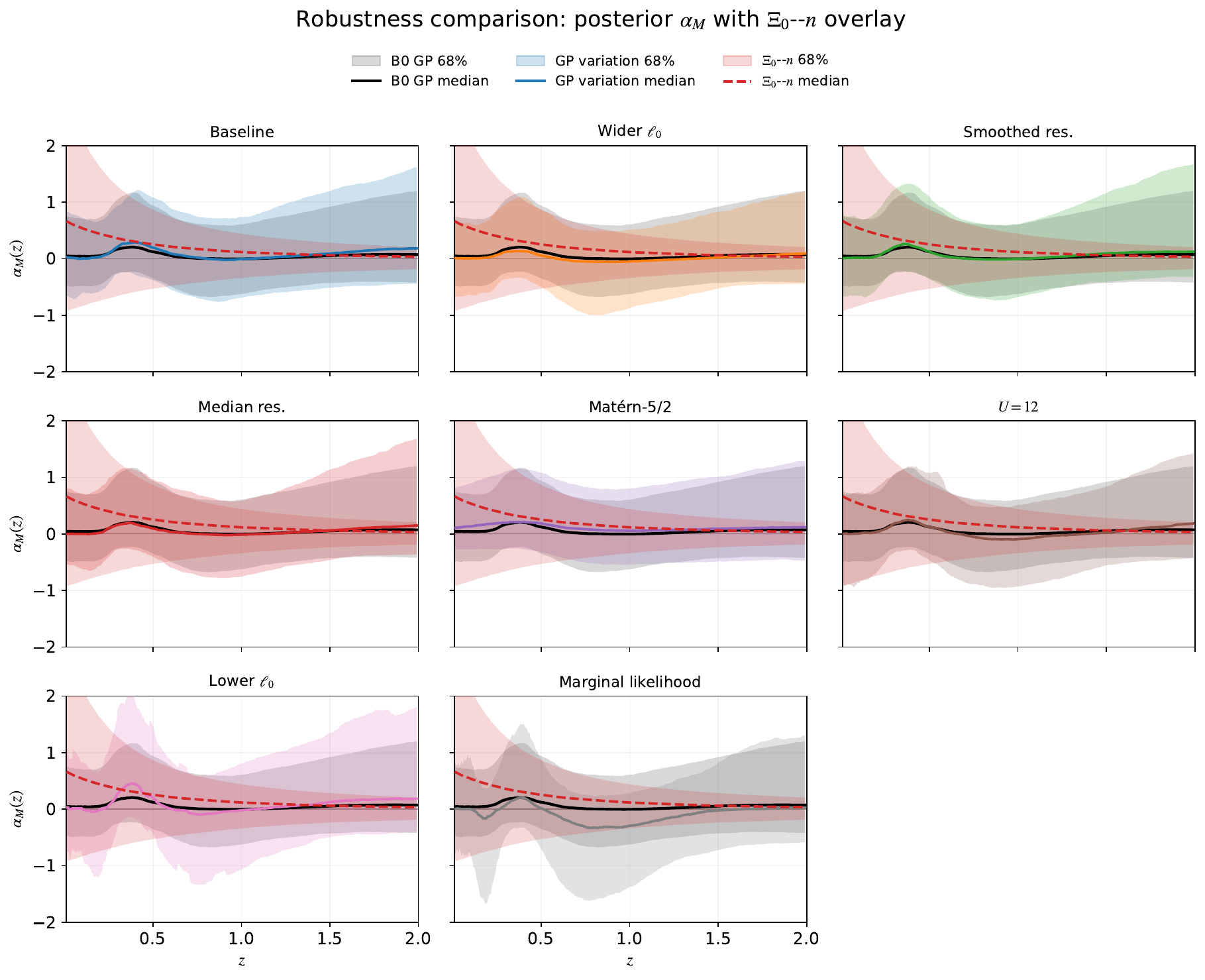}
  \caption{
  Robustness of the reconstructed GW friction function $\alpha_M(z)$ under
  GP modelling variations. The plotting convention is the same as in
  Fig.~\ref{fig:robustness_Xi_GWTC5}: gray denotes the calibrated baseline
  GP reconstruction, colored bands and curves denote the GP variant, and red
  denotes the $\Xi_0$--$n$ posterior. The local quantity $\alpha_M(z)$ is
  more sensitive than $\Xi(z)$ to smoothness and likelihood choices, but the
  main conclusion remains unchanged: the reconstructions are compatible with
  GR and with one another over the event-supported redshift range.
  }
  \label{fig:robustness_alphaM_GWTC5}
\end{figure*}

Figures~\ref{fig:robustness_Xi_GWTC5} and
\ref{fig:robustness_alphaM_GWTC5} show the result of this scan.
In each
case we compare the variant to the calibrated baseline reconstruction and to
the $\Xi_0$--$n$ posterior. 
The
distance-ratio reconstruction is the most stable quantity. Across all
variants, the GP median for $\Xi(z)$ remains close to the calibrated
baseline and close to GR, with differences that are small compared to the
posterior uncertainty. The largest changes appear in variants that broaden
or locally loosen the GP prior, such as the lower-$\ell_0$ and $U=12$ runs,
but even in these cases the posterior remains consistent with the baseline
and with $\Xi=1$.

The friction function $\alpha_M(z)$ is more sensitive to these choices, as
expected for a local derivative-like quantity. Variants that allow shorter
correlation scales or modify the likelihood treatment can produce more
localized structure in $\alpha_M(z)$, especially at low and intermediate
redshift. However, all variations remain compatible with GR. Furthermore, these features do not translate into a qualitatively
different accumulated distance ratio. 
This is important: the quantity most directly constrained by the luminosity-distance data is $\Xi(z)$, and its posterior is stable under the GP modelling variations.

The $\Xi_0$--$n$ posterior is shown as reference in the same panels. It is consistent with the GP reconstructions in all cases.
The robustness scan supports the interpretation
of the main result: the GWTC--5 data are consistent with GR, and this
conclusion is not driven by a specific GP kernel, resolution prescription, or
amplitude choice.

\section{Conclusions}\label{sect:conclusions}

In this work we have developed a non-parametric reconstruction of modified
gravitational-wave propagation based on Gaussian processes. Rather than
parametrizing directly the accumulated GW-to-EM luminosity-distance ratio
$\Xi(z)$, we reconstruct the local friction function $\alpha_M(z)$ and obtain
$\Xi(z)$ by integration. The construction is formulated so that the GW
luminosity distance remains monotonic and therefore invertible, while the
non-stationary GP kernel adapts its characteristic correlation scale to the
redshift resolution of the catalog.

We validated the method on simulated catalogs containing both a GR
fiducial and a non-trivial polexp propagation history that lies outside the
standard $\Xi_0$--$n$ functional form. In both cases the reconstruction
recovers the injected behavior within the posterior uncertainty, showing that
the GR-centered GP prior can accommodate significant and non-standard
departures when supported by sufficiently informative data.

We then applied the method to the GWTC--5 BBH catalog, jointly inferring the
BBH population, the modified-propagation history, and the background
cosmological parameters $H_0$, $\Omega_m$, and $w_0$. We find no evidence
for departures from GR: both $\alpha_M=0$ and $\Xi=1$ remain consistent with
the posterior over the redshift range probed by the catalog. This conclusion
is stable under variations of the GP kernel, the correlation-scale prior, the
resolution prescription, the GP amplitude prior, and the likelihood
treatment. In particular, while the local reconstruction of $\alpha_M(z)$ is
naturally more sensitive to these modelling choices, the accumulated
distance ratio $\Xi(z)$ is remarkably stable.

Our results complement the recent weakly modelled spectral-siren analysis of
Ref.~\cite{Colangeli:2026kbi}, which reconstructs the distance ratio
$\Xi(z)$ directly using a Fourier basis and likewise finds no evidence for
modified propagation in GWTC--5. The two approaches differ in emphasis.
Here the primary reconstructed quantity is the local friction
$\alpha_M(z)$, the background cosmology is varied jointly with the
propagation and population parameters, including $w_0$, and particular
attention is devoted to the dependence of the inference on the choice of
function-space prior and reconstruction settings.

More generally, an important aspect of this work is the development of
diagnostics that allow a flexible reconstruction to be compared with a
parametric model on a common footing. In particular, we have used:
\begin{itemize}
    \item a calibration of the GP amplitude prior through the integrated
    prior width of event-binned $\log\Xi$, so that the comparison with standard parametric models is not driven simply by different prior volumes;

    \item the corresponding prior-to-posterior contraction in the same function-space metric, which shows that, after this calibration, the GP and $\Xi_0$--$n$ reconstructions undergo different amounts of posterior contraction;
    
    \item a common function-space summary,
    $\Delta\langle\log\Xi\rangle_{\rm high-low}$, which makes it possible to
    compare propagation--population correlations even though the GP has no
    single parameter analogous to $\Xi_0$; in the limit where the low- and high-redshift averages reduce to $z=0$ and $z\to\infty$, respectively, this summary becomes $\log\Xi_0$ in the $\Xi_0$--$n$ parameterization.

    \item a principal-component analysis of the event-binned
    $\log\Xi$ covariance, which shows how the two models distribute posterior
    uncertainty across independent redshift-dependent modes.
\end{itemize}
These diagnostics reveal that the propagation--rate degeneracy remains
present in the flexible reconstruction. The difference is instead in the
geometry of the allowed propagation histories: the $\Xi_0$--$n$ posterior
is restricted to a single coherent mode, whereas the GP retains
additional independent directions in function space. We expect this type of
comparison to be useful more generally when assessing whether constraints
obtained with a fixed functional ansatz are driven by the data or by the
structure imposed by the parameterization.

The applicability of the present analysis is explicitly
restricted to luminal, massless, source-free tensor propagation, with
anomalous friction as its only modification and standard GR binary waveform
generation. This restriction belongs to the adopted likelihood: the standard spectral-siren mass conversion
does not hold for generic homogeneous time-dependent $c_T(z)$. 
Functional flexibility in $\Xi(z)$ alone therefore does not extend the
analysis to those theories.
This restriction applies to any constraint on the GW-to-EM distance ratio obtained with current spectral-siren techniques.
A nonluminal extension requires joint modelling
of mass redshifting and amplitude propagation. This would provide a possible next step, even though deviations from luminality are disfavoured, at least at low redhsift, by the stringent GW170817 constraint. Within the adopted luminal sector, our reconstruction remains much more general than parametric approaches in the redshift dependence of
physical GW friction.

A further extension would be to relax the assumption that
$d_L^{\rm GW}(z)$ is monotonic. Monotonicity is required here because the
likelihood is evaluated through the inverse map $z(d_L^{\rm GW})$. More
general likelihood constructions can avoid this inverse map by drawing
proposal samples directly in redshift space, as in
Ref.~\cite{Agapito:2026pbb}, and could therefore in principle accommodate
non-monotonic distance--redshift relations. The treatment of selection
effects hovever becomes less straightforward. Standard injection campaigns
provide finite sets of signals selected according to their detectability in
detector-frame parameter space, where luminosity distance rather than
redshift determines detectability.
They are therefore not immediately adapted to a likelihood formulated
entirely through redshift-space proposals when the distance--redshift map is
not invertible. A possible route would be to evaluate the detection
probability continuously and on the fly for proposed sources, using
selection-function emulators such as that of
Ref.~\cite{Callister:2024qyq}. Such an extension would permit substantially
more general propagation histories, but would require a dedicated likelihood
and selection-function design as well as significantly greater computational
development.

The increasing size and precision of GW catalogs will make these aspects
progressively more important. 
Flexible reconstructions provide a way to test GW propagation
without committing to a specific redshift dependence, but their interpretation
requires explicit control of the function-space prior and quantitative
diagnostics of how the posterior departs from and contracts relative to that
prior. In this work we have introduced and applied several concrete tools for
this purpose. Such prior--posterior checks are essential for
distinguishing information supplied by the data from structure imposed by a
non-parametric reconstruction, while retaining physical interpretability.

\normalsize

\section*{Data availability}

The population inference code used for this work is available at~Ref.~\cite{pymcpop}. 
The version used for this paper will be archived on Zenodo.
Data products for events in the GWTC-5.0 catalog are available at~\cite{ligo_scientific_collaboration_and_virgo_2025_17014085, ligo_scientific_collaboration_and_virgo_2023_8177023, ligo_scientific_collaboration_and_virgo_2022_6513631}.
Sensitivity estimates are available at~\cite{ligo_scientific_collaboration_2026_19500052}.

\let\oldaddcontentsline\addcontentsline
\renewcommand{\addcontentsline}[3]{}
\begin{acknowledgments}

We thank the LVK Tests of $\Lambda$CDM group members, Simone Mastrogiovanni, and Gr\'egoire Pierra for useful discussions. 
We also thank Elena Colangeli for internal LVK review.
This work received support from the French government under the France 2030 investment plan, as part of the Initiative d'Excellence d'Aix-Marseille Universit\'e -- A*MIDEX AMX-22-CEI-02. 
We acknowledge use of computational resources of the MesoCentre of Aix-Marseille University through the allocation b435-GWCROSS.\\

{\small This research has made use of data or software obtained from the Gravitational Wave Open Science Center (gwosc.org), a service of the LIGO Scientific Collaboration, the Virgo Collaboration, and KAGRA. This material is based upon work supported by NSF's LIGO Laboratory which is a major facility fully funded by the National Science Foundation, as well as the Science and Technology Facilities Council (STFC) of the United Kingdom, the Max-Planck-Society (MPS), and the State of Niedersachsen/Germany for support of the construction of Advanced LIGO and construction and operation of the GEO600 detector. Additional support for Advanced LIGO was provided by the Australian Research Council. Virgo is funded, through the European Gravitational Observatory (EGO), by the French Centre National de Recherche Scientifique (CNRS), the Italian Istituto Nazionale di Fisica Nucleare (INFN) and the Dutch Nikhef, with contributions by institutions from Belgium, Germany, Greece, Hungary, Ireland, Japan, Monaco, Poland, Portugal, Spain. KAGRA is supported by Ministry of Education, Culture, Sports, Science and Technology (MEXT), Japan Society for the Promotion of Science (JSPS) in Japan; National Research Foundation (NRF) and Ministry of Science and ICT (MSIT) in Korea; Academia Sinica (AS) and National Science and Technology Council (NSTC) in Taiwan.
}

\end{acknowledgments}
\let\addcontentsline\oldaddcontentsline

\let\oldaddcontentsline\addcontentsline
\renewcommand{\addcontentsline}[3]{}
\bibliography{bibliography}

\begin{thebibliography}{76}%
\makeatletter
\providecommand \@ifxundefined [1]{%
 \@ifx{#1\undefined}
}%
\providecommand \@ifnum [1]{%
 \ifnum #1\expandafter \@firstoftwo
 \else \expandafter \@secondoftwo
 \fi
}%
\providecommand \@ifx [1]{%
 \ifx #1\expandafter \@firstoftwo
 \else \expandafter \@secondoftwo
 \fi
}%
\providecommand \natexlab [1]{#1}%
\providecommand \enquote  [1]{``#1''}%
\providecommand \bibnamefont  [1]{#1}%
\providecommand \bibfnamefont [1]{#1}%
\providecommand \citenamefont [1]{#1}%
\providecommand \href@noop [0]{\@secondoftwo}%
\providecommand \href [0]{\begingroup \@sanitize@url \@href}%
\providecommand \@href[1]{\@@startlink{#1}\@@href}%
\providecommand \@@href[1]{\endgroup#1\@@endlink}%
\providecommand \@sanitize@url [0]{\catcode `\\12\catcode `\$12\catcode
  `\&12\catcode `\#12\catcode `\^12\catcode `\_12\catcode `\%12\relax}%
\providecommand \@@startlink[1]{}%
\providecommand \@@endlink[0]{}%
\providecommand \url  [0]{\begingroup\@sanitize@url \@url }%
\providecommand \@url [1]{\endgroup\@href {#1}{\urlprefix }}%
\providecommand \urlprefix  [0]{URL }%
\providecommand \Eprint [0]{\href }%
\providecommand \doibase [0]{https://doi.org/}%
\providecommand \selectlanguage [0]{\@gobble}%
\providecommand \bibinfo  [0]{\@secondoftwo}%
\providecommand \bibfield  [0]{\@secondoftwo}%
\providecommand \translation [1]{[#1]}%
\providecommand \BibitemOpen [0]{}%
\providecommand \bibitemStop [0]{}%
\providecommand \bibitemNoStop [0]{.\EOS\space}%
\providecommand \EOS [0]{\spacefactor3000\relax}%
\providecommand \BibitemShut  [1]{\csname bibitem#1\endcsname}%
\let\auto@bib@innerbib\@empty
\bibitem [{\citenamefont {Schutz}(1986)}]{Schutz:1986gp}%
  \BibitemOpen
  \bibfield  {author} {\bibinfo {author} {\bibfnamefont {B.~F.}\ \bibnamefont
  {Schutz}},\ }\href {https://doi.org/10.1038/323310a0} {\bibfield  {journal}
  {\bibinfo  {journal} {Nature}\ }\textbf {\bibinfo {volume} {323}},\ \bibinfo
  {pages} {310} (\bibinfo {year} {1986})}\BibitemShut {NoStop}%
\bibitem [{\citenamefont {Holz}\ and\ \citenamefont
  {Hughes}(2005)}]{Holz:2005df}%
  \BibitemOpen
  \bibfield  {author} {\bibinfo {author} {\bibfnamefont {D.~E.}\ \bibnamefont
  {Holz}}\ and\ \bibinfo {author} {\bibfnamefont {S.~A.}\ \bibnamefont
  {Hughes}},\ }\href {https://doi.org/10.1086/431341} {\bibfield  {journal}
  {\bibinfo  {journal} {Astrophys. J.}\ }\textbf {\bibinfo {volume} {629}},\
  \bibinfo {pages} {15} (\bibinfo {year} {2005})},\ \Eprint
  {https://arxiv.org/abs/astro-ph/0504616} {arXiv:astro-ph/0504616 [astro-ph]}
  \BibitemShut {NoStop}%
\bibitem [{\citenamefont {Taylor}\ \emph {et~al.}(2012)\citenamefont {Taylor},
  \citenamefont {Gair},\ and\ \citenamefont {Mandel}}]{Taylor:2011fs}%
  \BibitemOpen
  \bibfield  {author} {\bibinfo {author} {\bibfnamefont {S.~R.}\ \bibnamefont
  {Taylor}}, \bibinfo {author} {\bibfnamefont {J.~R.}\ \bibnamefont {Gair}},\
  and\ \bibinfo {author} {\bibfnamefont {I.}~\bibnamefont {Mandel}},\ }\href
  {https://doi.org/10.1103/PhysRevD.85.023535} {\bibfield  {journal} {\bibinfo
  {journal} {Phys. Rev.}\ }\textbf {\bibinfo {volume} {D85}},\ \bibinfo {pages}
  {023535} (\bibinfo {year} {2012})},\ \Eprint
  {https://arxiv.org/abs/1108.5161} {arXiv:1108.5161 [gr-qc]} \BibitemShut
  {NoStop}%
\bibitem [{\citenamefont {Farr}\ \emph {et~al.}(2019)\citenamefont {Farr},
  \citenamefont {Fishbach}, \citenamefont {Ye},\ and\ \citenamefont
  {Holz}}]{Farr:2019twy}%
  \BibitemOpen
  \bibfield  {author} {\bibinfo {author} {\bibfnamefont {W.~M.}\ \bibnamefont
  {Farr}}, \bibinfo {author} {\bibfnamefont {M.}~\bibnamefont {Fishbach}},
  \bibinfo {author} {\bibfnamefont {J.}~\bibnamefont {Ye}},\ and\ \bibinfo
  {author} {\bibfnamefont {D.}~\bibnamefont {Holz}},\ }\href
  {https://doi.org/10.3847/2041-8213/ab4284} {\bibfield  {journal} {\bibinfo
  {journal} {Astrophys. J. Lett.}\ }\textbf {\bibinfo {volume} {883}},\
  \bibinfo {pages} {L42} (\bibinfo {year} {2019})},\ \Eprint
  {https://arxiv.org/abs/1908.09084} {arXiv:1908.09084 [astro-ph.CO]}
  \BibitemShut {NoStop}%
\bibitem [{\citenamefont {Mastrogiovanni}\ \emph {et~al.}(2021)\citenamefont
  {Mastrogiovanni}, \citenamefont {Leyde}, \citenamefont {Karathanasis},
  \citenamefont {Chassande-Mottin}, \citenamefont {Steer}, \citenamefont
  {Gair}, \citenamefont {Ghosh}, \citenamefont {Gray}, \citenamefont
  {Mukherjee},\ and\ \citenamefont {Rinaldi}}]{Mastrogiovanni:2021wsd}%
  \BibitemOpen
  \bibfield  {author} {\bibinfo {author} {\bibfnamefont {S.}~\bibnamefont
  {Mastrogiovanni}}, \bibinfo {author} {\bibfnamefont {K.}~\bibnamefont
  {Leyde}}, \bibinfo {author} {\bibfnamefont {C.}~\bibnamefont {Karathanasis}},
  \bibinfo {author} {\bibfnamefont {E.}~\bibnamefont {Chassande-Mottin}},
  \bibinfo {author} {\bibfnamefont {D.~A.}\ \bibnamefont {Steer}}, \bibinfo
  {author} {\bibfnamefont {J.}~\bibnamefont {Gair}}, \bibinfo {author}
  {\bibfnamefont {A.}~\bibnamefont {Ghosh}}, \bibinfo {author} {\bibfnamefont
  {R.}~\bibnamefont {Gray}}, \bibinfo {author} {\bibfnamefont {S.}~\bibnamefont
  {Mukherjee}},\ and\ \bibinfo {author} {\bibfnamefont {S.}~\bibnamefont
  {Rinaldi}},\ }\href {https://doi.org/10.1103/PhysRevD.104.062009} {\bibfield
  {journal} {\bibinfo  {journal} {Phys. Rev. D}\ }\textbf {\bibinfo {volume}
  {104}},\ \bibinfo {pages} {062009} (\bibinfo {year} {2021})},\ \Eprint
  {https://arxiv.org/abs/2103.14663} {arXiv:2103.14663 [gr-qc]} \BibitemShut
  {NoStop}%
\bibitem [{\citenamefont {Ezquiaga}\ and\ \citenamefont
  {Holz}(2022)}]{Ezquiaga:2022zkx}%
  \BibitemOpen
  \bibfield  {author} {\bibinfo {author} {\bibfnamefont {J.~M.}\ \bibnamefont
  {Ezquiaga}}\ and\ \bibinfo {author} {\bibfnamefont {D.~E.}\ \bibnamefont
  {Holz}},\ }\href {https://doi.org/10.1103/PhysRevLett.129.061102} {\bibfield
  {journal} {\bibinfo  {journal} {Phys. Rev. Lett.}\ }\textbf {\bibinfo
  {volume} {129}},\ \bibinfo {pages} {061102} (\bibinfo {year} {2022})},\
  \Eprint {https://arxiv.org/abs/2202.08240} {arXiv:2202.08240 [astro-ph.CO]}
  \BibitemShut {NoStop}%
\bibitem [{\citenamefont {Mastrogiovanni}\ \emph {et~al.}(2023)\citenamefont
  {Mastrogiovanni}, \citenamefont {Laghi}, \citenamefont {Gray}, \citenamefont
  {Santoro}, \citenamefont {Ghosh}, \citenamefont {Karathanasis}, \citenamefont
  {Leyde}, \citenamefont {Steer}, \citenamefont {Perries},\ and\ \citenamefont
  {Pierra}}]{Mastrogiovanni:2023emh}%
  \BibitemOpen
  \bibfield  {author} {\bibinfo {author} {\bibfnamefont {S.}~\bibnamefont
  {Mastrogiovanni}}, \bibinfo {author} {\bibfnamefont {D.}~\bibnamefont
  {Laghi}}, \bibinfo {author} {\bibfnamefont {R.}~\bibnamefont {Gray}},
  \bibinfo {author} {\bibfnamefont {G.~C.}\ \bibnamefont {Santoro}}, \bibinfo
  {author} {\bibfnamefont {A.}~\bibnamefont {Ghosh}}, \bibinfo {author}
  {\bibfnamefont {C.}~\bibnamefont {Karathanasis}}, \bibinfo {author}
  {\bibfnamefont {K.}~\bibnamefont {Leyde}}, \bibinfo {author} {\bibfnamefont
  {D.~A.}\ \bibnamefont {Steer}}, \bibinfo {author} {\bibfnamefont
  {S.}~\bibnamefont {Perries}},\ and\ \bibinfo {author} {\bibfnamefont
  {G.}~\bibnamefont {Pierra}},\ }\href
  {https://doi.org/10.1103/PhysRevD.108.042002} {\bibfield  {journal} {\bibinfo
   {journal} {Phys. Rev. D}\ }\textbf {\bibinfo {volume} {108}},\ \bibinfo
  {pages} {042002} (\bibinfo {year} {2023})},\ \Eprint
  {https://arxiv.org/abs/2305.10488} {arXiv:2305.10488 [astro-ph.CO]}
  \BibitemShut {NoStop}%
\bibitem [{\citenamefont {Gray}\ \emph {et~al.}(2023)\citenamefont {Gray} \emph
  {et~al.}}]{Gray:2023wgj}%
  \BibitemOpen
  \bibfield  {author} {\bibinfo {author} {\bibfnamefont {R.}~\bibnamefont
  {Gray}} \emph {et~al.},\ }\href
  {https://doi.org/10.1088/1475-7516/2023/12/023} {\bibfield  {journal}
  {\bibinfo  {journal} {JCAP}\ }\textbf {\bibinfo {volume} {12}},\ \bibinfo
  {pages} {023}},\ \Eprint {https://arxiv.org/abs/2308.02281} {arXiv:2308.02281
  [astro-ph.CO]} \BibitemShut {NoStop}%
\bibitem [{\citenamefont {Borghi}\ \emph {et~al.}(2024)\citenamefont {Borghi},
  \citenamefont {Mancarella}, \citenamefont {Moresco}, \citenamefont
  {Tagliazucchi}, \citenamefont {Iacovelli}, \citenamefont {Cimatti},\ and\
  \citenamefont {Maggiore}}]{Borghi:2023opd}%
  \BibitemOpen
  \bibfield  {author} {\bibinfo {author} {\bibfnamefont {N.}~\bibnamefont
  {Borghi}}, \bibinfo {author} {\bibfnamefont {M.}~\bibnamefont {Mancarella}},
  \bibinfo {author} {\bibfnamefont {M.}~\bibnamefont {Moresco}}, \bibinfo
  {author} {\bibfnamefont {M.}~\bibnamefont {Tagliazucchi}}, \bibinfo {author}
  {\bibfnamefont {F.}~\bibnamefont {Iacovelli}}, \bibinfo {author}
  {\bibfnamefont {A.}~\bibnamefont {Cimatti}},\ and\ \bibinfo {author}
  {\bibfnamefont {M.}~\bibnamefont {Maggiore}},\ }\href
  {https://doi.org/10.3847/1538-4357/ad20eb} {\bibfield  {journal} {\bibinfo
  {journal} {Astrophys. J.}\ }\textbf {\bibinfo {volume} {964}},\ \bibinfo
  {pages} {191} (\bibinfo {year} {2024})},\ \Eprint
  {https://arxiv.org/abs/2312.05302} {arXiv:2312.05302 [astro-ph.CO]}
  \BibitemShut {NoStop}%
\bibitem [{\citenamefont {Saltas}\ \emph {et~al.}(2014)\citenamefont {Saltas},
  \citenamefont {Sawicki}, \citenamefont {Amendola},\ and\ \citenamefont
  {Kunz}}]{Saltas:2014dha}%
  \BibitemOpen
  \bibfield  {author} {\bibinfo {author} {\bibfnamefont {I.~D.}\ \bibnamefont
  {Saltas}}, \bibinfo {author} {\bibfnamefont {I.}~\bibnamefont {Sawicki}},
  \bibinfo {author} {\bibfnamefont {L.}~\bibnamefont {Amendola}},\ and\
  \bibinfo {author} {\bibfnamefont {M.}~\bibnamefont {Kunz}},\ }\href
  {https://doi.org/10.1103/PhysRevLett.113.191101} {\bibfield  {journal}
  {\bibinfo  {journal} {Phys. Rev. Lett.}\ }\textbf {\bibinfo {volume} {113}},\
  \bibinfo {pages} {191101} (\bibinfo {year} {2014})},\ \Eprint
  {https://arxiv.org/abs/1406.7139} {arXiv:1406.7139 [astro-ph.CO]}
  \BibitemShut {NoStop}%
\bibitem [{\citenamefont {Pettorino}\ and\ \citenamefont
  {Amendola}(2015)}]{Pettorino:2014bka}%
  \BibitemOpen
  \bibfield  {author} {\bibinfo {author} {\bibfnamefont {V.}~\bibnamefont
  {Pettorino}}\ and\ \bibinfo {author} {\bibfnamefont {L.}~\bibnamefont
  {Amendola}},\ }\href {https://doi.org/10.1016/j.physletb.2015.02.007}
  {\bibfield  {journal} {\bibinfo  {journal} {Phys. Lett. B}\ }\textbf
  {\bibinfo {volume} {742}},\ \bibinfo {pages} {353} (\bibinfo {year}
  {2015})},\ \Eprint {https://arxiv.org/abs/1408.2224} {arXiv:1408.2224
  [astro-ph.CO]} \BibitemShut {NoStop}%
\bibitem [{\citenamefont {Nishizawa}(2018)}]{Nishizawa:2017nef}%
  \BibitemOpen
  \bibfield  {author} {\bibinfo {author} {\bibfnamefont {A.}~\bibnamefont
  {Nishizawa}},\ }\href {https://doi.org/10.1103/PhysRevD.97.104037} {\bibfield
   {journal} {\bibinfo  {journal} {Phys. Rev. D}\ }\textbf {\bibinfo {volume}
  {97}},\ \bibinfo {pages} {104037} (\bibinfo {year} {2018})},\ \Eprint
  {https://arxiv.org/abs/1710.04825} {arXiv:1710.04825 [gr-qc]} \BibitemShut
  {NoStop}%
\bibitem [{\citenamefont {Amendola}\ \emph {et~al.}(2018)\citenamefont
  {Amendola}, \citenamefont {Sawicki}, \citenamefont {Kunz},\ and\
  \citenamefont {Saltas}}]{Amendola:2017ovw}%
  \BibitemOpen
  \bibfield  {author} {\bibinfo {author} {\bibfnamefont {L.}~\bibnamefont
  {Amendola}}, \bibinfo {author} {\bibfnamefont {I.}~\bibnamefont {Sawicki}},
  \bibinfo {author} {\bibfnamefont {M.}~\bibnamefont {Kunz}},\ and\ \bibinfo
  {author} {\bibfnamefont {I.~D.}\ \bibnamefont {Saltas}},\ }\href
  {https://doi.org/10.1088/1475-7516/2018/08/030} {\bibfield  {journal}
  {\bibinfo  {journal} {JCAP}\ }\textbf {\bibinfo {volume} {08}},\ \bibinfo
  {pages} {030}},\ \Eprint {https://arxiv.org/abs/1712.08623} {arXiv:1712.08623
  [astro-ph.CO]} \BibitemShut {NoStop}%
\bibitem [{\citenamefont {Belgacem}\ \emph
  {et~al.}(2018{\natexlab{a}})\citenamefont {Belgacem}, \citenamefont {Dirian},
  \citenamefont {Foffa},\ and\ \citenamefont {Maggiore}}]{Belgacem:2017ihm}%
  \BibitemOpen
  \bibfield  {author} {\bibinfo {author} {\bibfnamefont {E.}~\bibnamefont
  {Belgacem}}, \bibinfo {author} {\bibfnamefont {Y.}~\bibnamefont {Dirian}},
  \bibinfo {author} {\bibfnamefont {S.}~\bibnamefont {Foffa}},\ and\ \bibinfo
  {author} {\bibfnamefont {M.}~\bibnamefont {Maggiore}},\ }\href
  {https://doi.org/10.1103/PhysRevD.97.104066} {\bibfield  {journal} {\bibinfo
  {journal} {Phys. Rev. D}\ }\textbf {\bibinfo {volume} {97}},\ \bibinfo
  {pages} {104066} (\bibinfo {year} {2018}{\natexlab{a}})},\ \Eprint
  {https://arxiv.org/abs/1712.08108} {arXiv:1712.08108 [astro-ph.CO]}
  \BibitemShut {NoStop}%
\bibitem [{\citenamefont {Belgacem}\ \emph
  {et~al.}(2018{\natexlab{b}})\citenamefont {Belgacem}, \citenamefont {Dirian},
  \citenamefont {Foffa},\ and\ \citenamefont {Maggiore}}]{Belgacem:2018lbp}%
  \BibitemOpen
  \bibfield  {author} {\bibinfo {author} {\bibfnamefont {E.}~\bibnamefont
  {Belgacem}}, \bibinfo {author} {\bibfnamefont {Y.}~\bibnamefont {Dirian}},
  \bibinfo {author} {\bibfnamefont {S.}~\bibnamefont {Foffa}},\ and\ \bibinfo
  {author} {\bibfnamefont {M.}~\bibnamefont {Maggiore}},\ }\href
  {https://doi.org/10.1103/PhysRevD.98.023510} {\bibfield  {journal} {\bibinfo
  {journal} {Phys. Rev.}\ }\textbf {\bibinfo {volume} {D98}},\ \bibinfo {pages}
  {023510} (\bibinfo {year} {2018}{\natexlab{b}})},\ \Eprint
  {https://arxiv.org/abs/1805.08731} {arXiv:1805.08731 [gr-qc]} \BibitemShut
  {NoStop}%
\bibitem [{\citenamefont {Lagos}\ \emph {et~al.}(2019)\citenamefont {Lagos},
  \citenamefont {Fishbach}, \citenamefont {Landry},\ and\ \citenamefont
  {Holz}}]{Lagos:2019kds}%
  \BibitemOpen
  \bibfield  {author} {\bibinfo {author} {\bibfnamefont {M.}~\bibnamefont
  {Lagos}}, \bibinfo {author} {\bibfnamefont {M.}~\bibnamefont {Fishbach}},
  \bibinfo {author} {\bibfnamefont {P.}~\bibnamefont {Landry}},\ and\ \bibinfo
  {author} {\bibfnamefont {D.~E.}\ \bibnamefont {Holz}},\ }\href
  {https://doi.org/10.1103/PhysRevD.99.083504} {\bibfield  {journal} {\bibinfo
  {journal} {Phys. Rev. D}\ }\textbf {\bibinfo {volume} {99}},\ \bibinfo
  {pages} {083504} (\bibinfo {year} {2019})},\ \Eprint
  {https://arxiv.org/abs/1901.03321} {arXiv:1901.03321 [astro-ph.CO]}
  \BibitemShut {NoStop}%
\bibitem [{\citenamefont {Ishak}\ \emph {et~al.}(2025)\citenamefont {Ishak}
  \emph {et~al.}}]{Ishak:2024jhs}%
  \BibitemOpen
  \bibfield  {author} {\bibinfo {author} {\bibfnamefont {M.}~\bibnamefont
  {Ishak}} \emph {et~al.},\ }\href
  {https://doi.org/10.1088/1475-7516/2025/09/053} {\bibfield  {journal}
  {\bibinfo  {journal} {JCAP}\ }\textbf {\bibinfo {volume} {09}},\ \bibinfo
  {pages} {053}},\ \Eprint {https://arxiv.org/abs/2411.12026} {arXiv:2411.12026
  [astro-ph.CO]} \BibitemShut {NoStop}%
\bibitem [{\citenamefont {Belgacem}\ \emph {et~al.}(2019)\citenamefont
  {Belgacem} \emph {et~al.}}]{LISACosmologyWorkingGroup:2019mwx}%
  \BibitemOpen
  \bibfield  {author} {\bibinfo {author} {\bibfnamefont {E.}~\bibnamefont
  {Belgacem}} \emph {et~al.} (\bibinfo {collaboration} {LISA Cosmology Working
  Group}),\ }\href {https://doi.org/10.1088/1475-7516/2019/07/024} {\bibfield
  {journal} {\bibinfo  {journal} {JCAP}\ }\textbf {\bibinfo {volume} {07}},\
  \bibinfo {pages} {024}},\ \Eprint {https://arxiv.org/abs/1906.01593}
  {arXiv:1906.01593 [astro-ph.CO]} \BibitemShut {NoStop}%
\bibitem [{\citenamefont {Finke}\ \emph {et~al.}(2021)\citenamefont {Finke},
  \citenamefont {Foffa}, \citenamefont {Iacovelli}, \citenamefont {Maggiore},\
  and\ \citenamefont {Mancarella}}]{Finke:2021aom}%
  \BibitemOpen
  \bibfield  {author} {\bibinfo {author} {\bibfnamefont {A.}~\bibnamefont
  {Finke}}, \bibinfo {author} {\bibfnamefont {S.}~\bibnamefont {Foffa}},
  \bibinfo {author} {\bibfnamefont {F.}~\bibnamefont {Iacovelli}}, \bibinfo
  {author} {\bibfnamefont {M.}~\bibnamefont {Maggiore}},\ and\ \bibinfo
  {author} {\bibfnamefont {M.}~\bibnamefont {Mancarella}},\ }\href
  {https://doi.org/10.1088/1475-7516/2021/08/026} {\bibfield  {journal}
  {\bibinfo  {journal} {JCAP}\ }\textbf {\bibinfo {volume} {08}},\ \bibinfo
  {pages} {026}},\ \Eprint {https://arxiv.org/abs/2101.12660} {arXiv:2101.12660
  [astro-ph.CO]} \BibitemShut {NoStop}%
\bibitem [{\citenamefont {Ezquiaga}(2021)}]{Ezquiaga:2021ayr}%
  \BibitemOpen
  \bibfield  {author} {\bibinfo {author} {\bibfnamefont {J.~M.}\ \bibnamefont
  {Ezquiaga}},\ }\href {https://doi.org/10.1016/j.physletb.2021.136665}
  {\bibfield  {journal} {\bibinfo  {journal} {Phys. Lett. B}\ }\textbf
  {\bibinfo {volume} {822}},\ \bibinfo {pages} {136665} (\bibinfo {year}
  {2021})},\ \Eprint {https://arxiv.org/abs/2104.05139} {arXiv:2104.05139
  [astro-ph.CO]} \BibitemShut {NoStop}%
\bibitem [{\citenamefont {Mancarella}\ \emph {et~al.}(2022)\citenamefont
  {Mancarella}, \citenamefont {Genoud-Prachex},\ and\ \citenamefont
  {Maggiore}}]{Mancarella:2021ecn}%
  \BibitemOpen
  \bibfield  {author} {\bibinfo {author} {\bibfnamefont {M.}~\bibnamefont
  {Mancarella}}, \bibinfo {author} {\bibfnamefont {E.}~\bibnamefont
  {Genoud-Prachex}},\ and\ \bibinfo {author} {\bibfnamefont {M.}~\bibnamefont
  {Maggiore}},\ }\href {https://doi.org/10.1103/PhysRevD.105.064030} {\bibfield
   {journal} {\bibinfo  {journal} {Phys. Rev. D}\ }\textbf {\bibinfo {volume}
  {105}},\ \bibinfo {pages} {064030} (\bibinfo {year} {2022})},\ \Eprint
  {https://arxiv.org/abs/2112.05728} {arXiv:2112.05728 [gr-qc]} \BibitemShut
  {NoStop}%
\bibitem [{\citenamefont {Leyde}\ \emph {et~al.}(2022)\citenamefont {Leyde},
  \citenamefont {Mastrogiovanni}, \citenamefont {Steer}, \citenamefont
  {Chassande-Mottin},\ and\ \citenamefont {Karathanasis}}]{Leyde:2022orh}%
  \BibitemOpen
  \bibfield  {author} {\bibinfo {author} {\bibfnamefont {K.}~\bibnamefont
  {Leyde}}, \bibinfo {author} {\bibfnamefont {S.}~\bibnamefont
  {Mastrogiovanni}}, \bibinfo {author} {\bibfnamefont {D.~A.}\ \bibnamefont
  {Steer}}, \bibinfo {author} {\bibfnamefont {E.}~\bibnamefont
  {Chassande-Mottin}},\ and\ \bibinfo {author} {\bibfnamefont {C.}~\bibnamefont
  {Karathanasis}},\ }\href {https://doi.org/10.1088/1475-7516/2022/09/012}
  {\bibfield  {journal} {\bibinfo  {journal} {JCAP}\ }\textbf {\bibinfo
  {volume} {09}},\ \bibinfo {pages} {012}},\ \Eprint
  {https://arxiv.org/abs/2202.00025} {arXiv:2202.00025 [gr-qc]} \BibitemShut
  {NoStop}%
\bibitem [{\citenamefont {Chen}\ \emph {et~al.}(2024)\citenamefont {Chen},
  \citenamefont {Gray},\ and\ \citenamefont {Baker}}]{Chen:2023wpj}%
  \BibitemOpen
  \bibfield  {author} {\bibinfo {author} {\bibfnamefont {A.}~\bibnamefont
  {Chen}}, \bibinfo {author} {\bibfnamefont {R.}~\bibnamefont {Gray}},\ and\
  \bibinfo {author} {\bibfnamefont {T.}~\bibnamefont {Baker}},\ }\href
  {https://doi.org/10.1088/1475-7516/2024/02/035} {\bibfield  {journal}
  {\bibinfo  {journal} {JCAP}\ }\textbf {\bibinfo {volume} {02}},\ \bibinfo
  {pages} {035}},\ \Eprint {https://arxiv.org/abs/2309.03833} {arXiv:2309.03833
  [gr-qc]} \BibitemShut {NoStop}%
\bibitem [{\citenamefont {Abac}\ \emph
  {et~al.}(2026{\natexlab{a}})\citenamefont {Abac} \emph
  {et~al.}}]{LIGOScientific:2025jau}%
  \BibitemOpen
  \bibfield  {author} {\bibinfo {author} {\bibfnamefont {A.~G.}\ \bibnamefont
  {Abac}} \emph {et~al.} (\bibinfo {collaboration} {LIGO Scientific, Virgo,
  KAGRA}),\ }\href {https://doi.org/10.3847/2041-8213/ae69dd} {\bibfield
  {journal} {\bibinfo  {journal} {Astrophys. J. Lett.}\ }\textbf {\bibinfo
  {volume} {1007}},\ \bibinfo {pages} {L17} (\bibinfo {year}
  {2026}{\natexlab{a}})},\ \Eprint {https://arxiv.org/abs/2509.04348}
  {arXiv:2509.04348 [astro-ph.CO]} \BibitemShut {NoStop}%
\bibitem [{\citenamefont {Langlois}\ and\ \citenamefont
  {Noui}(2016)}]{Langlois:2015cwa}%
  \BibitemOpen
  \bibfield  {author} {\bibinfo {author} {\bibfnamefont {D.}~\bibnamefont
  {Langlois}}\ and\ \bibinfo {author} {\bibfnamefont {K.}~\bibnamefont
  {Noui}},\ }\href {https://doi.org/10.1088/1475-7516/2016/02/034} {\bibfield
  {journal} {\bibinfo  {journal} {JCAP}\ }\textbf {\bibinfo {volume} {02}},\
  \bibinfo {pages} {034}},\ \Eprint {https://arxiv.org/abs/1510.06930}
  {arXiv:1510.06930 [gr-qc]} \BibitemShut {NoStop}%
\bibitem [{\citenamefont {Dvali}\ \emph {et~al.}(2000)\citenamefont {Dvali},
  \citenamefont {Gabadadze},\ and\ \citenamefont {Porrati}}]{Dvali:2000hr}%
  \BibitemOpen
  \bibfield  {author} {\bibinfo {author} {\bibfnamefont {G.~R.}\ \bibnamefont
  {Dvali}}, \bibinfo {author} {\bibfnamefont {G.}~\bibnamefont {Gabadadze}},\
  and\ \bibinfo {author} {\bibfnamefont {M.}~\bibnamefont {Porrati}},\ }\href
  {https://doi.org/10.1016/S0370-2693(00)00669-9} {\bibfield  {journal}
  {\bibinfo  {journal} {Phys. Lett. B}\ }\textbf {\bibinfo {volume} {485}},\
  \bibinfo {pages} {208} (\bibinfo {year} {2000})},\ \Eprint
  {https://arxiv.org/abs/hep-th/0005016} {arXiv:hep-th/0005016} \BibitemShut
  {NoStop}%
\bibitem [{\citenamefont {Hassan}\ and\ \citenamefont
  {Rosen}(2012)}]{Hassan:2011zd}%
  \BibitemOpen
  \bibfield  {author} {\bibinfo {author} {\bibfnamefont {S.~F.}\ \bibnamefont
  {Hassan}}\ and\ \bibinfo {author} {\bibfnamefont {R.~A.}\ \bibnamefont
  {Rosen}},\ }\href {https://doi.org/10.1007/JHEP02(2012)126} {\bibfield
  {journal} {\bibinfo  {journal} {JHEP}\ }\textbf {\bibinfo {volume} {02}},\
  \bibinfo {pages} {126}},\ \Eprint {https://arxiv.org/abs/1109.3515}
  {arXiv:1109.3515 [hep-th]} \BibitemShut {NoStop}%
\bibitem [{\citenamefont {Bellini}\ and\ \citenamefont
  {Sawicki}(2014)}]{Bellini:2014fua}%
  \BibitemOpen
  \bibfield  {author} {\bibinfo {author} {\bibfnamefont {E.}~\bibnamefont
  {Bellini}}\ and\ \bibinfo {author} {\bibfnamefont {I.}~\bibnamefont
  {Sawicki}},\ }\href {https://doi.org/10.1088/1475-7516/2014/07/050}
  {\bibfield  {journal} {\bibinfo  {journal} {JCAP}\ }\textbf {\bibinfo
  {volume} {07}},\ \bibinfo {pages} {050}},\ \Eprint
  {https://arxiv.org/abs/1404.3713} {arXiv:1404.3713 [astro-ph.CO]}
  \BibitemShut {NoStop}%
\bibitem [{\citenamefont {Gleyzes}\ \emph {et~al.}(2015)\citenamefont
  {Gleyzes}, \citenamefont {Langlois},\ and\ \citenamefont
  {Vernizzi}}]{Gleyzes:2014rba}%
  \BibitemOpen
  \bibfield  {author} {\bibinfo {author} {\bibfnamefont {J.}~\bibnamefont
  {Gleyzes}}, \bibinfo {author} {\bibfnamefont {D.}~\bibnamefont {Langlois}},\
  and\ \bibinfo {author} {\bibfnamefont {F.}~\bibnamefont {Vernizzi}},\ }\href
  {https://doi.org/10.1142/S021827181443010X} {\bibfield  {journal} {\bibinfo
  {journal} {Int. J. Mod. Phys. D}\ }\textbf {\bibinfo {volume} {23}},\
  \bibinfo {pages} {1443010} (\bibinfo {year} {2015})},\ \Eprint
  {https://arxiv.org/abs/1411.3712} {arXiv:1411.3712 [hep-th]} \BibitemShut
  {NoStop}%
\bibitem [{\citenamefont {Noller}\ and\ \citenamefont
  {Nicola}(2019)}]{Noller:2018wyv}%
  \BibitemOpen
  \bibfield  {author} {\bibinfo {author} {\bibfnamefont {J.}~\bibnamefont
  {Noller}}\ and\ \bibinfo {author} {\bibfnamefont {A.}~\bibnamefont
  {Nicola}},\ }\href {https://doi.org/10.1103/PhysRevD.99.103502} {\bibfield
  {journal} {\bibinfo  {journal} {Phys. Rev. D}\ }\textbf {\bibinfo {volume}
  {99}},\ \bibinfo {pages} {103502} (\bibinfo {year} {2019})},\ \Eprint
  {https://arxiv.org/abs/1811.12928} {arXiv:1811.12928 [astro-ph.CO]}
  \BibitemShut {NoStop}%
\bibitem [{\citenamefont {Baker}\ and\ \citenamefont
  {Harrison}(2021)}]{Baker:2020apq}%
  \BibitemOpen
  \bibfield  {author} {\bibinfo {author} {\bibfnamefont {T.}~\bibnamefont
  {Baker}}\ and\ \bibinfo {author} {\bibfnamefont {I.}~\bibnamefont
  {Harrison}},\ }\href {https://doi.org/10.1088/1475-7516/2021/01/068}
  {\bibfield  {journal} {\bibinfo  {journal} {JCAP}\ }\textbf {\bibinfo
  {volume} {01}},\ \bibinfo {pages} {068}},\ \Eprint
  {https://arxiv.org/abs/2007.13791} {arXiv:2007.13791 [astro-ph.CO]}
  \BibitemShut {NoStop}%
\bibitem [{\citenamefont {Seraille}\ \emph {et~al.}(2024)\citenamefont
  {Seraille}, \citenamefont {Noller},\ and\ \citenamefont
  {Sherwin}}]{Seraille:2024beb}%
  \BibitemOpen
  \bibfield  {author} {\bibinfo {author} {\bibfnamefont {E.}~\bibnamefont
  {Seraille}}, \bibinfo {author} {\bibfnamefont {J.}~\bibnamefont {Noller}},\
  and\ \bibinfo {author} {\bibfnamefont {B.~D.}\ \bibnamefont {Sherwin}},\
  }\href {https://doi.org/10.1103/PhysRevD.110.123525} {\bibfield  {journal}
  {\bibinfo  {journal} {Phys. Rev. D}\ }\textbf {\bibinfo {volume} {110}},\
  \bibinfo {pages} {123525} (\bibinfo {year} {2024})},\ \Eprint
  {https://arxiv.org/abs/2401.06221} {arXiv:2401.06221 [astro-ph.CO]}
  \BibitemShut {NoStop}%
\bibitem [{LIG(2026{\natexlab{a}})}]{LIGOScientific:2026uyd}%
  \BibitemOpen
  \href@noop {} {\  (\bibinfo {year} {2026}{\natexlab{a}})},\ \Eprint
  {https://arxiv.org/abs/2605.27227} {arXiv:2605.27227 [astro-ph.CO]}
  \BibitemShut {NoStop}%
\bibitem [{\citenamefont {Linder}\ \emph {et~al.}(2016)\citenamefont {Linder},
  \citenamefont {Seng\"or},\ and\ \citenamefont {Watson}}]{Linder:2015rcz}%
  \BibitemOpen
  \bibfield  {author} {\bibinfo {author} {\bibfnamefont {E.~V.}\ \bibnamefont
  {Linder}}, \bibinfo {author} {\bibfnamefont {G.}~\bibnamefont {Seng\"or}},\
  and\ \bibinfo {author} {\bibfnamefont {S.}~\bibnamefont {Watson}},\ }\href
  {https://doi.org/10.1088/1475-7516/2016/05/053} {\bibfield  {journal}
  {\bibinfo  {journal} {JCAP}\ }\textbf {\bibinfo {volume} {05}},\ \bibinfo
  {pages} {053}},\ \Eprint {https://arxiv.org/abs/1512.06180} {arXiv:1512.06180
  [astro-ph.CO]} \BibitemShut {NoStop}%
\bibitem [{\citenamefont {Linder}(2017)}]{Linder:2016wqw}%
  \BibitemOpen
  \bibfield  {author} {\bibinfo {author} {\bibfnamefont {E.~V.}\ \bibnamefont
  {Linder}},\ }\href {https://doi.org/10.1103/PhysRevD.95.023518} {\bibfield
  {journal} {\bibinfo  {journal} {Phys. Rev. D}\ }\textbf {\bibinfo {volume}
  {95}},\ \bibinfo {pages} {023518} (\bibinfo {year} {2017})},\ \Eprint
  {https://arxiv.org/abs/1607.03113} {arXiv:1607.03113 [astro-ph.CO]}
  \BibitemShut {NoStop}%
\bibitem [{\citenamefont {Denissenya}\ and\ \citenamefont
  {Linder}(2018)}]{Denissenya:2018mqs}%
  \BibitemOpen
  \bibfield  {author} {\bibinfo {author} {\bibfnamefont {M.}~\bibnamefont
  {Denissenya}}\ and\ \bibinfo {author} {\bibfnamefont {E.~V.}\ \bibnamefont
  {Linder}},\ }\href {https://doi.org/10.1088/1475-7516/2018/11/010} {\bibfield
   {journal} {\bibinfo  {journal} {JCAP}\ }\textbf {\bibinfo {volume} {11}},\
  \bibinfo {pages} {010}},\ \Eprint {https://arxiv.org/abs/1808.00013}
  {arXiv:1808.00013 [astro-ph.CO]} \BibitemShut {NoStop}%
\bibitem [{\citenamefont {Gleyzes}(2017)}]{Gleyzes:2017kpi}%
  \BibitemOpen
  \bibfield  {author} {\bibinfo {author} {\bibfnamefont {J.}~\bibnamefont
  {Gleyzes}},\ }\href {https://doi.org/10.1103/PhysRevD.96.063516} {\bibfield
  {journal} {\bibinfo  {journal} {Phys. Rev. D}\ }\textbf {\bibinfo {volume}
  {96}},\ \bibinfo {pages} {063516} (\bibinfo {year} {2017})},\ \Eprint
  {https://arxiv.org/abs/1705.04714} {arXiv:1705.04714 [astro-ph.CO]}
  \BibitemShut {NoStop}%
\bibitem [{\citenamefont {Pierra}\ \emph {et~al.}(2026)\citenamefont {Pierra},
  \citenamefont {Colombo},\ and\ \citenamefont
  {Mastrogiovanni}}]{Pierra:2025hoc}%
  \BibitemOpen
  \bibfield  {author} {\bibinfo {author} {\bibfnamefont {G.}~\bibnamefont
  {Pierra}}, \bibinfo {author} {\bibfnamefont {A.}~\bibnamefont {Colombo}},\
  and\ \bibinfo {author} {\bibfnamefont {S.}~\bibnamefont {Mastrogiovanni}},\
  }\href {https://doi.org/10.1088/1361-6382/ae849b} {\bibfield  {journal}
  {\bibinfo  {journal} {Class. Quant. Grav.}\ }\textbf {\bibinfo {volume}
  {43}},\ \bibinfo {pages} {155015} (\bibinfo {year} {2026})},\ \Eprint
  {https://arxiv.org/abs/2511.11795} {arXiv:2511.11795 [astro-ph.CO]}
  \BibitemShut {NoStop}%
\bibitem [{\citenamefont {Belgacem}\ \emph {et~al.}(2020)\citenamefont
  {Belgacem}, \citenamefont {Foffa}, \citenamefont {Maggiore},\ and\
  \citenamefont {Yang}}]{Belgacem:2019zzu}%
  \BibitemOpen
  \bibfield  {author} {\bibinfo {author} {\bibfnamefont {E.}~\bibnamefont
  {Belgacem}}, \bibinfo {author} {\bibfnamefont {S.}~\bibnamefont {Foffa}},
  \bibinfo {author} {\bibfnamefont {M.}~\bibnamefont {Maggiore}},\ and\
  \bibinfo {author} {\bibfnamefont {T.}~\bibnamefont {Yang}},\ }\href
  {https://doi.org/10.1103/PhysRevD.101.063505} {\bibfield  {journal} {\bibinfo
   {journal} {Phys. Rev. D}\ }\textbf {\bibinfo {volume} {101}},\ \bibinfo
  {pages} {063505} (\bibinfo {year} {2020})},\ \Eprint
  {https://arxiv.org/abs/1911.11497} {arXiv:1911.11497 [astro-ph.CO]}
  \BibitemShut {NoStop}%
\bibitem [{\citenamefont {Colangeli}\ \emph {et~al.}(2026)\citenamefont
  {Colangeli}, \citenamefont {Leyde}, \citenamefont {Baker},\ and\
  \citenamefont {Chen}}]{Colangeli:2026kbi}%
  \BibitemOpen
  \bibfield  {author} {\bibinfo {author} {\bibfnamefont {E.}~\bibnamefont
  {Colangeli}}, \bibinfo {author} {\bibfnamefont {K.}~\bibnamefont {Leyde}},
  \bibinfo {author} {\bibfnamefont {T.}~\bibnamefont {Baker}},\ and\ \bibinfo
  {author} {\bibfnamefont {A.}~\bibnamefont {Chen}},\ }\href@noop {} {\
  (\bibinfo {year} {2026})},\ \Eprint {https://arxiv.org/abs/2608.18310}
  {arXiv:2608.18310 [gr-qc]} \BibitemShut {NoStop}%
\bibitem [{\citenamefont {Abac}\ \emph
  {et~al.}(2026{\natexlab{b}})\citenamefont {Abac} \emph
  {et~al.}}]{LIGOScientific:2026sit}%
  \BibitemOpen
  \bibfield  {author} {\bibinfo {author} {\bibfnamefont {N.}~\bibnamefont
  {Abac}} \emph {et~al.} (\bibinfo {collaboration} {LIGO Scientific, VIRGO,
  KAGRA}),\ }\href@noop {} {\  (\bibinfo {year} {2026}{\natexlab{b}})},\
  \Eprint {https://arxiv.org/abs/2605.27223} {arXiv:2605.27223 [gr-qc]}
  \BibitemShut {NoStop}%
\bibitem [{\citenamefont {Ezquiaga}\ and\ \citenamefont
  {Zumalac\'arregui}(2018)}]{Ezquiaga:2018btd}%
  \BibitemOpen
  \bibfield  {author} {\bibinfo {author} {\bibfnamefont {J.~M.}\ \bibnamefont
  {Ezquiaga}}\ and\ \bibinfo {author} {\bibfnamefont {M.}~\bibnamefont
  {Zumalac\'arregui}},\ }\href {https://doi.org/10.3389/fspas.2018.00044}
  {\bibfield  {journal} {\bibinfo  {journal} {Front. Astron. Space Sci.}\
  }\textbf {\bibinfo {volume} {5}},\ \bibinfo {pages} {44} (\bibinfo {year}
  {2018})},\ \Eprint {https://arxiv.org/abs/1807.09241} {arXiv:1807.09241
  [astro-ph.CO]} \BibitemShut {NoStop}%
\bibitem [{\citenamefont {Abbott}\ \emph
  {et~al.}(2017{\natexlab{a}})\citenamefont {Abbott} \emph
  {et~al.}}]{TheLIGOScientific:2017qsa}%
  \BibitemOpen
  \bibfield  {author} {\bibinfo {author} {\bibfnamefont {B.~P.}\ \bibnamefont
  {Abbott}} \emph {et~al.},\ }\href
  {https://doi.org/10.1103/PhysRevLett.119.161101} {\bibfield  {journal}
  {\bibinfo  {journal} {Phys. Rev. Lett.}\ }\textbf {\bibinfo {volume} {119}},\
  \bibinfo {pages} {161101} (\bibinfo {year} {2017}{\natexlab{a}})},\ \Eprint
  {https://arxiv.org/abs/1710.05832} {arXiv:1710.05832 [gr-qc]} \BibitemShut
  {NoStop}%
\bibitem [{\citenamefont {Abbott}\ \emph
  {et~al.}(2017{\natexlab{b}})\citenamefont {Abbott} \emph
  {et~al.}}]{LIGOScientific:2017zic}%
  \BibitemOpen
  \bibfield  {author} {\bibinfo {author} {\bibfnamefont {B.~P.}\ \bibnamefont
  {Abbott}} \emph {et~al.} (\bibinfo {collaboration} {LIGO Scientific, Virgo,
  Fermi-GBM, INTEGRAL}),\ }\href {https://doi.org/10.3847/2041-8213/aa920c}
  {\bibfield  {journal} {\bibinfo  {journal} {Astrophys. J. Lett.}\ }\textbf
  {\bibinfo {volume} {848}},\ \bibinfo {pages} {L13} (\bibinfo {year}
  {2017}{\natexlab{b}})},\ \Eprint {https://arxiv.org/abs/1710.05834}
  {arXiv:1710.05834 [astro-ph.HE]} \BibitemShut {NoStop}%
\bibitem [{\citenamefont {Simpson}\ \emph {et~al.}(2017)\citenamefont
  {Simpson}, \citenamefont {Rue}, \citenamefont {Riebler}, \citenamefont
  {Martins},\ and\ \citenamefont {S{\o}rbye}}]{Simpson:2017pc}%
  \BibitemOpen
  \bibfield  {author} {\bibinfo {author} {\bibfnamefont {D.}~\bibnamefont
  {Simpson}}, \bibinfo {author} {\bibfnamefont {H.}~\bibnamefont {Rue}},
  \bibinfo {author} {\bibfnamefont {A.}~\bibnamefont {Riebler}}, \bibinfo
  {author} {\bibfnamefont {T.~G.}\ \bibnamefont {Martins}},\ and\ \bibinfo
  {author} {\bibfnamefont {S.~H.}\ \bibnamefont {S{\o}rbye}},\ }\href
  {https://doi.org/10.1214/16-STS576} {\bibfield  {journal} {\bibinfo
  {journal} {Statist. Sci.}\ }\textbf {\bibinfo {volume} {32}},\ \bibinfo
  {pages} {1} (\bibinfo {year} {2017})},\ \Eprint
  {https://arxiv.org/abs/1403.4630} {arXiv:1403.4630 [stat.ME]} \BibitemShut
  {NoStop}%
\bibitem [{\citenamefont {Fuglstad}\ \emph {et~al.}(2019)\citenamefont
  {Fuglstad}, \citenamefont {Simpson}, \citenamefont {Lindgren},\ and\
  \citenamefont {Rue}}]{Fuglstad:2018pc}%
  \BibitemOpen
  \bibfield  {author} {\bibinfo {author} {\bibfnamefont {G.-A.}\ \bibnamefont
  {Fuglstad}}, \bibinfo {author} {\bibfnamefont {D.}~\bibnamefont {Simpson}},
  \bibinfo {author} {\bibfnamefont {F.}~\bibnamefont {Lindgren}},\ and\
  \bibinfo {author} {\bibfnamefont {H.}~\bibnamefont {Rue}},\ }\href
  {https://doi.org/10.1080/01621459.2017.1415907} {\bibfield  {journal}
  {\bibinfo  {journal} {J. Amer. Statist. Assoc.}\ }\textbf {\bibinfo {volume}
  {114}},\ \bibinfo {pages} {445} (\bibinfo {year} {2019})},\ \Eprint
  {https://arxiv.org/abs/1503.00256} {arXiv:1503.00256 [stat.ME]} \BibitemShut
  {NoStop}%
\bibitem [{\citenamefont {Abac}\ \emph {et~al.}(2025)\citenamefont {Abac} \emph
  {et~al.}}]{LIGOScientific:2025pvj}%
  \BibitemOpen
  \bibfield  {author} {\bibinfo {author} {\bibfnamefont {A.~G.}\ \bibnamefont
  {Abac}} \emph {et~al.} (\bibinfo {collaboration} {LIGO Scientific, VIRGO,
  KAGRA}),\ }\href@noop {} {\  (\bibinfo {year} {2025})},\ \Eprint
  {https://arxiv.org/abs/2508.18083} {arXiv:2508.18083 [astro-ph.HE]}
  \BibitemShut {NoStop}%
\bibitem [{LIG(2026{\natexlab{b}})}]{LIGOScientific:2026ctl}%
  \BibitemOpen
  \href@noop {} {\  (\bibinfo {year} {2026}{\natexlab{b}})},\ \Eprint
  {https://arxiv.org/abs/2605.27226} {arXiv:2605.27226 [astro-ph.HE]}
  \BibitemShut {NoStop}%
\bibitem [{\citenamefont {Agapito}\ \emph {et~al.}(2026)\citenamefont
  {Agapito}, \citenamefont {De~Renzis},\ and\ \citenamefont
  {Mancarella}}]{Agapito:2026kak}%
  \BibitemOpen
  \bibfield  {author} {\bibinfo {author} {\bibfnamefont {A.}~\bibnamefont
  {Agapito}}, \bibinfo {author} {\bibfnamefont {V.}~\bibnamefont {De~Renzis}},\
  and\ \bibinfo {author} {\bibfnamefont {M.}~\bibnamefont {Mancarella}},\
  }\href@noop {} {\  (\bibinfo {year} {2026})},\ \Eprint
  {https://arxiv.org/abs/2605.20112} {arXiv:2605.20112 [astro-ph.CO]}
  \BibitemShut {NoStop}%
\bibitem [{\citenamefont {Mandel}\ \emph {et~al.}(2019)\citenamefont {Mandel},
  \citenamefont {Farr},\ and\ \citenamefont {Gair}}]{Mandel:2018mve}%
  \BibitemOpen
  \bibfield  {author} {\bibinfo {author} {\bibfnamefont {I.}~\bibnamefont
  {Mandel}}, \bibinfo {author} {\bibfnamefont {W.~M.}\ \bibnamefont {Farr}},\
  and\ \bibinfo {author} {\bibfnamefont {J.~R.}\ \bibnamefont {Gair}},\ }\href
  {https://doi.org/10.1093/mnras/stz896} {\bibfield  {journal} {\bibinfo
  {journal} {Mon. Not. R. Astron. Soc.}\ }\textbf {\bibinfo {volume} {486}},\
  \bibinfo {pages} {1086} (\bibinfo {year} {2019})},\ \Eprint
  {https://arxiv.org/abs/1809.02063} {arXiv:1809.02063 [physics.data-an]}
  \BibitemShut {NoStop}%
\bibitem [{\citenamefont {Vitale}\ \emph {et~al.}(2020)\citenamefont {Vitale},
  \citenamefont {Gerosa}, \citenamefont {Farr},\ and\ \citenamefont
  {Taylor}}]{Vitale:2020aaz}%
  \BibitemOpen
  \bibfield  {author} {\bibinfo {author} {\bibfnamefont {S.}~\bibnamefont
  {Vitale}}, \bibinfo {author} {\bibfnamefont {D.}~\bibnamefont {Gerosa}},
  \bibinfo {author} {\bibfnamefont {W.~M.}\ \bibnamefont {Farr}},\ and\
  \bibinfo {author} {\bibfnamefont {S.~R.}\ \bibnamefont {Taylor}}\ }\href
  {https://doi.org/10.1007/978-981-15-4702-7\_45-1}
  {10.1007/978-981-15-4702-7\_45-1} (\bibinfo {year} {2020}),\ \Eprint
  {https://arxiv.org/abs/2007.05579} {arXiv:2007.05579 [astro-ph.IM]}
  \BibitemShut {NoStop}%
\bibitem [{\citenamefont {Tiwari}(2018)}]{Tiwari:2017ndi}%
  \BibitemOpen
  \bibfield  {author} {\bibinfo {author} {\bibfnamefont {V.}~\bibnamefont
  {Tiwari}},\ }\href {https://doi.org/10.1088/1361-6382/aac89d} {\bibfield
  {journal} {\bibinfo  {journal} {Class. Quantum Grav.}\ }\textbf {\bibinfo
  {volume} {35}},\ \bibinfo {pages} {145009} (\bibinfo {year} {2018})},\
  \Eprint {https://arxiv.org/abs/1712.00482} {arXiv:1712.00482 [astro-ph.HE]}
  \BibitemShut {NoStop}%
\bibitem [{\citenamefont {Farr}(2019)}]{Farr:2019rap}%
  \BibitemOpen
  \bibfield  {author} {\bibinfo {author} {\bibfnamefont {W.~M.}\ \bibnamefont
  {Farr}},\ }\href {https://doi.org/10.3847/2515-5172/ab1d5f} {\bibfield
  {journal} {\bibinfo  {journal} {Research Notes of the AAS}\ }\textbf
  {\bibinfo {volume} {3}},\ \bibinfo {pages} {66} (\bibinfo {year} {2019})},\
  \Eprint {https://arxiv.org/abs/1904.10879} {arXiv:1904.10879 [astro-ph.IM]}
  \BibitemShut {NoStop}%
\bibitem [{\citenamefont {Talbot}\ and\ \citenamefont
  {Golomb}(2023)}]{Talbot:2023pex}%
  \BibitemOpen
  \bibfield  {author} {\bibinfo {author} {\bibfnamefont {C.}~\bibnamefont
  {Talbot}}\ and\ \bibinfo {author} {\bibfnamefont {J.}~\bibnamefont
  {Golomb}},\ }\href {https://doi.org/10.1093/mnras/stad2968} {\bibfield
  {journal} {\bibinfo  {journal} {Mon. Not. Roy. Astron. Soc.}\ }\textbf
  {\bibinfo {volume} {526}},\ \bibinfo {pages} {3495} (\bibinfo {year}
  {2023})},\ \Eprint {https://arxiv.org/abs/2304.06138} {arXiv:2304.06138
  [astro-ph.IM]} \BibitemShut {NoStop}%
\bibitem [{\citenamefont {Heinzel}\ and\ \citenamefont
  {Vitale}(2025)}]{Heinzel:2025ogf}%
  \BibitemOpen
  \bibfield  {author} {\bibinfo {author} {\bibfnamefont {J.}~\bibnamefont
  {Heinzel}}\ and\ \bibinfo {author} {\bibfnamefont {S.}~\bibnamefont
  {Vitale}},\ }\href@noop {} {\  (\bibinfo {year} {2025})},\ \Eprint
  {https://arxiv.org/abs/2509.07221} {arXiv:2509.07221 [astro-ph.HE]}
  \BibitemShut {NoStop}%
\bibitem [{\citenamefont {Mancarella}(2025)}]{pymcpop}%
  \BibitemOpen
  \bibfield  {author} {\bibinfo {author} {\bibfnamefont {M.}~\bibnamefont
  {Mancarella}},\ }\href@noop {} {}\bibinfo {howpublished}
  {\href{https://github.com/CosmoStatGW/pymcpop-gw}{github.com/CosmoStatGW/pymcpop-gw}}
  (\bibinfo {year} {2025})\BibitemShut {NoStop}%
\bibitem [{\citenamefont {Mancarella}\ and\ \citenamefont
  {Gerosa}(2025)}]{Mancarella:2025uat}%
  \BibitemOpen
  \bibfield  {author} {\bibinfo {author} {\bibfnamefont {M.}~\bibnamefont
  {Mancarella}}\ and\ \bibinfo {author} {\bibfnamefont {D.}~\bibnamefont
  {Gerosa}},\ }\href {https://doi.org/10.1103/PhysRevD.111.103012} {\bibfield
  {journal} {\bibinfo  {journal} {Phys. Rev. D}\ }\textbf {\bibinfo {volume}
  {111}},\ \bibinfo {pages} {103012} (\bibinfo {year} {2025})},\ \Eprint
  {https://arxiv.org/abs/2502.12156} {arXiv:2502.12156 [gr-qc]} \BibitemShut
  {NoStop}%
\bibitem [{\citenamefont {Hoffman}\ and\ \citenamefont
  {Gelman}(2014)}]{JMLR:v15:hoffman14a}%
  \BibitemOpen
  \bibfield  {author} {\bibinfo {author} {\bibfnamefont {M.~D.}\ \bibnamefont
  {Hoffman}}\ and\ \bibinfo {author} {\bibfnamefont {A.}~\bibnamefont
  {Gelman}},\ }\href {http://jmlr.org/papers/v15/hoffman14a.html} {\bibfield
  {journal} {\bibinfo  {journal} {J. Mach. Learn. Res.}\ }\textbf {\bibinfo
  {volume} {15}},\ \bibinfo {pages} {1593} (\bibinfo {year} {2014})},\ \Eprint
  {https://arxiv.org/abs/1111.4246} {arXiv:1111.4246 [stat.CO]} \BibitemShut
  {NoStop}%
\bibitem [{\citenamefont {Brooks}\ \emph {et~al.}(2011)\citenamefont {Brooks},
  \citenamefont {Gelman}, \citenamefont {Jones},\ and\ \citenamefont
  {Meng}}]{Brooks_2011}%
  \BibitemOpen
  \bibfield  {author} {\bibinfo {author} {\bibfnamefont {S.}~\bibnamefont
  {Brooks}}, \bibinfo {author} {\bibfnamefont {A.}~\bibnamefont {Gelman}},
  \bibinfo {author} {\bibfnamefont {G.}~\bibnamefont {Jones}},\ and\ \bibinfo
  {author} {\bibfnamefont {X.-L.}\ \bibnamefont {Meng}},\ }\href
  {https://doi.org/10.1201/b10905} {\emph {\bibinfo {title} {Handbook of Markov
  Chain Monte Carlo}}}\ (\bibinfo  {publisher} {Chapman and Hall/CRC},\
  \bibinfo {year} {2011})\BibitemShut {NoStop}%
\bibitem [{\citenamefont {Phan}\ \emph {et~al.}(2019)\citenamefont {Phan},
  \citenamefont {Pradhan},\ and\ \citenamefont
  {Jankowiak}}]{phan2019composable}%
  \BibitemOpen
  \bibfield  {author} {\bibinfo {author} {\bibfnamefont {D.}~\bibnamefont
  {Phan}}, \bibinfo {author} {\bibfnamefont {N.}~\bibnamefont {Pradhan}},\ and\
  \bibinfo {author} {\bibfnamefont {M.}~\bibnamefont {Jankowiak}},\ }\href@noop
  {} {\bibfield  {journal} {\bibinfo  {journal} {{NeurIPS 2019}}\ } (\bibinfo
  {year} {2019})},\ \Eprint {https://arxiv.org/abs/1912.11554}
  {arXiv:1912.11554 [stat]} \BibitemShut {NoStop}%
\bibitem [{\citenamefont {Bingham}\ \emph {et~al.}(2019)\citenamefont
  {Bingham}, \citenamefont {Chen}, \citenamefont {Jankowiak}, \citenamefont
  {Obermeyer}, \citenamefont {Pradhan}, \citenamefont {Karaletsos},
  \citenamefont {Singh}, \citenamefont {Szerlip}, \citenamefont {Horsfall},\
  and\ \citenamefont {Goodman}}]{bingham2019pyro}%
  \BibitemOpen
  \bibfield  {author} {\bibinfo {author} {\bibfnamefont {E.}~\bibnamefont
  {Bingham}}, \bibinfo {author} {\bibfnamefont {J.~P.}\ \bibnamefont {Chen}},
  \bibinfo {author} {\bibfnamefont {M.}~\bibnamefont {Jankowiak}}, \bibinfo
  {author} {\bibfnamefont {F.}~\bibnamefont {Obermeyer}}, \bibinfo {author}
  {\bibfnamefont {N.}~\bibnamefont {Pradhan}}, \bibinfo {author} {\bibfnamefont
  {T.}~\bibnamefont {Karaletsos}}, \bibinfo {author} {\bibfnamefont
  {R.}~\bibnamefont {Singh}}, \bibinfo {author} {\bibfnamefont {P.~A.}\
  \bibnamefont {Szerlip}}, \bibinfo {author} {\bibfnamefont {P.}~\bibnamefont
  {Horsfall}},\ and\ \bibinfo {author} {\bibfnamefont {N.~D.}\ \bibnamefont
  {Goodman}},\ }\href {http://jmlr.org/papers/v20/18-403.html} {\bibfield
  {journal} {\bibinfo  {journal} {J. Mach. Learn. Res.}\ }\textbf {\bibinfo
  {volume} {20}},\ \bibinfo {pages} {28:1} (\bibinfo {year} {2019})},\ \Eprint
  {https://arxiv.org/abs/1810.09538} {arXiv:1810.09538} \BibitemShut {NoStop}%
\bibitem [{\citenamefont {Gelman}\ and\ \citenamefont
  {Rubin}(1992)}]{10.1214/ss/1177011136}%
  \BibitemOpen
  \bibfield  {author} {\bibinfo {author} {\bibfnamefont {A.}~\bibnamefont
  {Gelman}}\ and\ \bibinfo {author} {\bibfnamefont {D.~B.}\ \bibnamefont
  {Rubin}},\ }\href {https://doi.org/10.1214/ss/1177011136} {\bibfield
  {journal} {\bibinfo  {journal} {Stat. Sci.}\ }\textbf {\bibinfo {volume}
  {7}},\ \bibinfo {pages} {457 } (\bibinfo {year} {1992})}\BibitemShut
  {NoStop}%
\bibitem [{\citenamefont {Farah}\ \emph {et~al.}(2023)\citenamefont {Farah},
  \citenamefont {Edelman}, \citenamefont {Zevin}, \citenamefont {Fishbach},
  \citenamefont {Ezquiaga}, \citenamefont {Farr},\ and\ \citenamefont
  {Holz}}]{Farah:2023vsc}%
  \BibitemOpen
  \bibfield  {author} {\bibinfo {author} {\bibfnamefont {A.~M.}\ \bibnamefont
  {Farah}}, \bibinfo {author} {\bibfnamefont {B.}~\bibnamefont {Edelman}},
  \bibinfo {author} {\bibfnamefont {M.}~\bibnamefont {Zevin}}, \bibinfo
  {author} {\bibfnamefont {M.}~\bibnamefont {Fishbach}}, \bibinfo {author}
  {\bibfnamefont {J.~M.}\ \bibnamefont {Ezquiaga}}, \bibinfo {author}
  {\bibfnamefont {B.}~\bibnamefont {Farr}},\ and\ \bibinfo {author}
  {\bibfnamefont {D.~E.}\ \bibnamefont {Holz}},\ }\href
  {https://doi.org/10.3847/1538-4357/aced02} {\bibfield  {journal} {\bibinfo
  {journal} {Astrophys. J.}\ }\textbf {\bibinfo {volume} {955}},\ \bibinfo
  {pages} {107} (\bibinfo {year} {2023})},\ \Eprint
  {https://arxiv.org/abs/2301.00834} {arXiv:2301.00834 [astro-ph.HE]}
  \BibitemShut {NoStop}%
\bibitem [{LIG(2026{\natexlab{c}})}]{LIGOScientific:2026wfs}%
  \BibitemOpen
  \href@noop {} {\  (\bibinfo {year} {2026}{\natexlab{c}})},\ \Eprint
  {https://arxiv.org/abs/2605.27225} {arXiv:2605.27225 [gr-qc]} \BibitemShut
  {NoStop}%
\bibitem [{\citenamefont {Collaboration}\ \emph {et~al.}(2025)\citenamefont
  {Collaboration}, \citenamefont {Collaboration},\ and\ \citenamefont
  {Collaboration}}]{ligo_scientific_collaboration_and_virgo_2025_17014085}%
  \BibitemOpen
  \bibfield  {author} {\bibinfo {author} {\bibfnamefont {L.~S.}\ \bibnamefont
  {Collaboration}}, \bibinfo {author} {\bibfnamefont {V.}~\bibnamefont
  {Collaboration}},\ and\ \bibinfo {author} {\bibfnamefont {K.}~\bibnamefont
  {Collaboration}},\ }\href {https://doi.org/10.5281/zenodo.17014085}
  {10.5281/zenodo.17014085} (\bibinfo {year} {2025})\BibitemShut {NoStop}%
\bibitem [{\citenamefont {Collaboration}\ \emph {et~al.}(2023)\citenamefont
  {Collaboration}, \citenamefont {Collaboration},\ and\ \citenamefont
  {Collaboration}}]{ligo_scientific_collaboration_and_virgo_2023_8177023}%
  \BibitemOpen
  \bibfield  {author} {\bibinfo {author} {\bibfnamefont {L.~S.}\ \bibnamefont
  {Collaboration}}, \bibinfo {author} {\bibfnamefont {V.}~\bibnamefont
  {Collaboration}},\ and\ \bibinfo {author} {\bibfnamefont {K.}~\bibnamefont
  {Collaboration}},\ }\href {https://doi.org/10.5281/zenodo.8177023}
  {10.5281/zenodo.8177023} (\bibinfo {year} {2023})\BibitemShut {NoStop}%
\bibitem [{\citenamefont {Collaboration}\ and\ \citenamefont
  {Collaboration}(2022)}]{ligo_scientific_collaboration_and_virgo_2022_6513631}%
  \BibitemOpen
  \bibfield  {author} {\bibinfo {author} {\bibfnamefont {L.~S.}\ \bibnamefont
  {Collaboration}}\ and\ \bibinfo {author} {\bibfnamefont {V.}~\bibnamefont
  {Collaboration}},\ }\href {https://doi.org/10.5281/zenodo.6513631}
  {10.5281/zenodo.6513631} (\bibinfo {year} {2022})\BibitemShut {NoStop}%
\bibitem [{\citenamefont {Collaboration}\ \emph
  {et~al.}(2026{\natexlab{a}})\citenamefont {Collaboration}, \citenamefont
  {Collaboration},\ and\ \citenamefont
  {Collaboration}}]{ligo_scientific_collaboration_and_virgo_2026_20348005}%
  \BibitemOpen
  \bibfield  {author} {\bibinfo {author} {\bibfnamefont {L.~S.}\ \bibnamefont
  {Collaboration}}, \bibinfo {author} {\bibfnamefont {V.}~\bibnamefont
  {Collaboration}},\ and\ \bibinfo {author} {\bibfnamefont {K.}~\bibnamefont
  {Collaboration}},\ }\href {https://doi.org/10.5281/zenodo.20348005}
  {10.5281/zenodo.20348005} (\bibinfo {year} {2026}{\natexlab{a}})\BibitemShut
  {NoStop}%
\bibitem [{\citenamefont {Collaboration}\ \emph
  {et~al.}(2026{\natexlab{b}})\citenamefont {Collaboration}, \citenamefont
  {Collaboration},\ and\ \citenamefont
  {Collaboration}}]{ligo_scientific_collaboration_2026_19500052}%
  \BibitemOpen
  \bibfield  {author} {\bibinfo {author} {\bibfnamefont {L.~S.}\ \bibnamefont
  {Collaboration}}, \bibinfo {author} {\bibfnamefont {V.}~\bibnamefont
  {Collaboration}},\ and\ \bibinfo {author} {\bibfnamefont {K.}~\bibnamefont
  {Collaboration}},\ }\href {https://doi.org/10.5281/zenodo.19500052}
  {10.5281/zenodo.19500052} (\bibinfo {year} {2026}{\natexlab{b}})\BibitemShut
  {NoStop}%
\bibitem [{\citenamefont {Essick}\ \emph {et~al.}(2025)\citenamefont {Essick}
  \emph {et~al.}}]{Essick:2025zed}%
  \BibitemOpen
  \bibfield  {author} {\bibinfo {author} {\bibfnamefont {R.}~\bibnamefont
  {Essick}} \emph {et~al.},\ }\href {https://doi.org/10.1103/44x3-hv3y}
  {\bibfield  {journal} {\bibinfo  {journal} {Phys. Rev. D}\ }\textbf {\bibinfo
  {volume} {112}},\ \bibinfo {pages} {102001} (\bibinfo {year} {2025})},\
  \Eprint {https://arxiv.org/abs/2508.10638} {arXiv:2508.10638 [gr-qc]}
  \BibitemShut {NoStop}%
\bibitem [{\citenamefont {Essick}(2023{\natexlab{a}})}]{Essick:2023toz}%
  \BibitemOpen
  \bibfield  {author} {\bibinfo {author} {\bibfnamefont {R.}~\bibnamefont
  {Essick}},\ }\href {https://doi.org/10.1103/PhysRevD.108.043011} {\bibfield
  {journal} {\bibinfo  {journal} {Phys. Rev. D}\ }\textbf {\bibinfo {volume}
  {108}},\ \bibinfo {pages} {043011} (\bibinfo {year} {2023}{\natexlab{a}})},\
  \Eprint {https://arxiv.org/abs/2307.02765} {arXiv:2307.02765 [gr-qc]}
  \BibitemShut {NoStop}%
\bibitem [{\citenamefont {Essick}(2023{\natexlab{b}})}]{PhysRevD.108.043011}%
  \BibitemOpen
  \bibfield  {author} {\bibinfo {author} {\bibfnamefont {R.}~\bibnamefont
  {Essick}},\ }\href {https://doi.org/10.1103/PhysRevD.108.043011} {\bibfield
  {journal} {\bibinfo  {journal} {Phys. Rev. D}\ }\textbf {\bibinfo {volume}
  {108}},\ \bibinfo {pages} {043011} (\bibinfo {year}
  {2023}{\natexlab{b}})}\BibitemShut {NoStop}%
\bibitem [{\citenamefont {Agapito}\ and\ \citenamefont
  {Mancarella}(2026)}]{Agapito:2026pbb}%
  \BibitemOpen
  \bibfield  {author} {\bibinfo {author} {\bibfnamefont {A.}~\bibnamefont
  {Agapito}}\ and\ \bibinfo {author} {\bibfnamefont {M.}~\bibnamefont
  {Mancarella}},\ }\href@noop {} {\  (\bibinfo {year} {2026})},\ \Eprint
  {https://arxiv.org/abs/2609.38110} {arXiv:2609.38110 [astro-ph.CO]}
  \BibitemShut {NoStop}%
\bibitem [{\citenamefont {Callister}\ \emph {et~al.}(2024)\citenamefont
  {Callister}, \citenamefont {Essick},\ and\ \citenamefont
  {Holz}}]{Callister:2024qyq}%
  \BibitemOpen
  \bibfield  {author} {\bibinfo {author} {\bibfnamefont {T.~A.}\ \bibnamefont
  {Callister}}, \bibinfo {author} {\bibfnamefont {R.}~\bibnamefont {Essick}},\
  and\ \bibinfo {author} {\bibfnamefont {D.~E.}\ \bibnamefont {Holz}},\ }\href
  {https://doi.org/10.1103/PhysRevD.110.123041} {\bibfield  {journal} {\bibinfo
   {journal} {Phys. Rev. D}\ }\textbf {\bibinfo {volume} {110}},\ \bibinfo
  {pages} {123041} (\bibinfo {year} {2024})},\ \Eprint
  {https://arxiv.org/abs/2408.16828} {arXiv:2408.16828 [astro-ph.HE]}
  \BibitemShut {NoStop}%
\bibitem [{\citenamefont {Tagliazucchi}\ \emph {et~al.}(2026)\citenamefont
  {Tagliazucchi}, \citenamefont {Moresco}, \citenamefont {Agapito},
  \citenamefont {Mancarella}, \citenamefont {Ferraiuolo}, \citenamefont
  {Mastrogiovanni}, \citenamefont {Borghi}, \citenamefont {Pannarale},\ and\
  \citenamefont {Bonacorsi}}]{Tagliazucchi:2026dpr}%
  \BibitemOpen
  \bibfield  {author} {\bibinfo {author} {\bibfnamefont {M.}~\bibnamefont
  {Tagliazucchi}}, \bibinfo {author} {\bibfnamefont {M.}~\bibnamefont
  {Moresco}}, \bibinfo {author} {\bibfnamefont {A.}~\bibnamefont {Agapito}},
  \bibinfo {author} {\bibfnamefont {M.}~\bibnamefont {Mancarella}}, \bibinfo
  {author} {\bibfnamefont {S.}~\bibnamefont {Ferraiuolo}}, \bibinfo {author}
  {\bibfnamefont {S.}~\bibnamefont {Mastrogiovanni}}, \bibinfo {author}
  {\bibfnamefont {N.}~\bibnamefont {Borghi}}, \bibinfo {author} {\bibfnamefont
  {F.}~\bibnamefont {Pannarale}},\ and\ \bibinfo {author} {\bibfnamefont
  {D.}~\bibnamefont {Bonacorsi}},\ }\href
  {https://doi.org/10.1051/0004-6361/202659517} {\bibfield  {journal} {\bibinfo
   {journal} {Astron. Astrophys.}\ }\textbf {\bibinfo {volume} {711}},\
  \bibinfo {pages} {A120} (\bibinfo {year} {2026})},\ \Eprint
  {https://arxiv.org/abs/2602.17756} {arXiv:2602.17756 [astro-ph.CO]}
  \BibitemShut {NoStop}%
\bibitem [{\citenamefont {Essick}\ and\ \citenamefont
  {Farr}(2022)}]{Essick:2022ojx}%
  \BibitemOpen
  \bibfield  {author} {\bibinfo {author} {\bibfnamefont {R.}~\bibnamefont
  {Essick}}\ and\ \bibinfo {author} {\bibfnamefont {W.}~\bibnamefont {Farr}},\
  }\href@noop {} {\  (\bibinfo {year} {2022})},\ \Eprint
  {https://arxiv.org/abs/2204.00461} {arXiv:2204.00461 [astro-ph.IM]}
  \BibitemShut {NoStop}%
\end{thebibliography}%
\let\addcontentsline\oldaddcontentsline

\appendix

\section{Population-model definitions}
\label{app:models_nonevol}

This appendix gives the explicit functional form of the population model. The model follows the default GWTC-4.0 BBH population prescription
of Ref.~\cite{LIGOScientific:2025pvj} for the primary-mass mixture structure, with the
sole difference that we apply the smoothstep low-mass taper defined in
Sec.~\ref{app:taper_nonevol}.

We write the source-population density as
\begin{equation}
p_{\rm pop}(\theta\mid\Lambda)
= p(z\mid\Lambda_z)\,
  p(m_1\mid\Lambda_m)\,
  p(m_2\mid m_1,\Lambda_m)\,
  p(\chi\mid\Lambda_\chi),
\end{equation}
where $\theta$ collects source-frame parameters (masses, spins, redshift), and
$\Lambda=\{\Lambda_{\rm cosmo},\Lambda_z,\Lambda_m,\Lambda_\chi\}$.

\paragraph{Primary-mass model: broken power law with two Gaussian peaks}
\label{app:dpldp_nonevol}

The primary-mass spectrum is a mixture of a broken
power law and two truncated Gaussians:
\begin{equation}
\begin{split}
p(m_1\mid\Lambda_m)
= &
\frac{T(m_1)}{\mathcal{Z}_{m_1}}
\Big[
\lambda_0\,p_{\rm BPL}(m_1)
+\lambda_1\,p_{{\rm G},1}(m_1) \\
&
+\lambda_2\,p_{{\rm G},2}(m_1)
\Big],
\end{split}
\end{equation}
defined on $m_1\in[m_{1,\min},m_{1,\max}]$, where $\lambda_i\ge 0$ and
$\lambda_0+\lambda_1+\lambda_2=1$. The normalization constant $\mathcal{Z}_{m_1}$ is
chosen such that $\int_{m_{1,\min}}^{m_{1,\max}} p(m_1\mid\Lambda_m)\,{\rm d}m_1=1$.

The broken power law has slopes $\alpha_1$ and $\alpha_2$ around a break mass $m_b$:
\begin{equation}
p_{\rm BPL}(m_1)\propto
\begin{cases}
\left(\dfrac{m_1}{m_b}\right)^{-\alpha_1}, & m_1<m_b,\\[6pt]
\left(\dfrac{m_1}{m_b}\right)^{-\alpha_2}, & m_1\ge m_b,
\end{cases}
\end{equation}
with support restricted to $[m_{1,\min},m_{1,\max}]$.
The Gaussian components are
\begin{equation}
p_{{\rm G},i}(m_1)\propto
\exp\!\left[-\frac{(m_1-\mu_i)^2}{2\sigma_i^2}\right],
\qquad i=1,2,
\end{equation}
each truncated to $[m_{1,\min},m_{1,\max}]$ and renormalized on that interval.

\paragraph{Secondary-mass model}
\label{app:m2_nonevol}

The secondary mass is modeled conditionally on $m_1$ as a power law in $m_2$
with low-mass taper and the physical constraint $m_2\le m_1$:
\begin{equation}
p(m_2\mid m_1,\Lambda_m)
=
\frac{1}{C(m_1)}
\,m_2^{\beta}\,T(m_2)\,\mathcal{H}(m_1-m_2),
\end{equation}
where $\mathcal{H}$ is the Heaviside step function and the normalization factor
\begin{equation}
C(m_1)=\int_{m_{2,\min}}^{m_1} m_2^{\beta}\,T(m_2)\,{\rm d}m_2
\end{equation}
ensures $\int p(m_2\mid m_1,\Lambda_m)\,{\rm d}m_2=1$ for each $m_1$.

\paragraph{Low-mass tapering (smoothstep)}
\label{app:taper_nonevol}

To avoid hard cutoffs and ensure differentiability, we apply a smoothstep taper at the
low-mass edge of both $m_1$ and $m_2$. For any mass variable $m$ with low-mass bound $m_{\min}$
and taper width $\delta m$, we define
\begin{equation}
t(m)=\frac{m-m_{\min}}{\delta m},
\qquad
T(m)=
\begin{cases}
0, & t\le 0,\\
3t^2-2t^3, & 0<t<1,\\
1, & t\ge 1.
\end{cases}
\end{equation}
We apply $T(m_1)$ with $(m_{1,\min},\delta m_1)$ and $T(m_2)$ with $(m_{2,\min},\delta m_2)$.
This smoothstep prescription differs from the low-mass smoothing used in the GWTC-4.0 population
analysis~\cite{LIGOScientific:2025pvj} but leaves the remainder of the model unchanged.

\paragraph{Auxiliary parametrization for low-mass bounds}
\label{app:bounds_nonevol}

The low-mass bound of the primary mass is sampled via an auxiliary variable $u\in(0,1)$:
\begin{equation}
m_{1,\min}=3 + (10-3)\,u^{3/2},
\qquad u\sim \mathrm{Uniform}(0,1),
\end{equation}
so that $m_{1,\min}\in[3,10]\,M_\odot$ while preserving a smooth unconstrained sampling geometry.
The secondary low-mass bound is constructed conditionally using $v\in(0,1)$:
\begin{equation}
m_{2,\min}=3 + v\,(m_{1,\min}-3),
\qquad v\sim \mathrm{Uniform}(0,1),
\end{equation}
which enforces $m_{2,\min}\le m_{1,\min}$ by construction.

\paragraph{Redshift distribution (Madau--Dickinson)}
\label{app:md_nonevol}

The population redshift distribution is
\begin{equation}
p(z\mid\Lambda_z)\propto
\frac{{\rm d}V_c}{{\rm d}z}\,
\frac{1}{1+z}\,
\psi(z\mid\gamma,\kappa,z_p),
\end{equation}
where ${\rm d}V_c/{\rm d}z$ is the differential comoving volume element in the assumed cosmology and
$(1+z)^{-1}$ accounts for time dilation. We use a Madau--Dickinson merger-rate density
\begin{equation}
\psi(z\mid\gamma,\kappa,z_p)
=
\frac{(1+z)^{\gamma}}
{1+\left(\dfrac{1+z}{1+z_p}\right)^{\gamma+\kappa}}.
\end{equation}

\section{Statistical framework}
\label{app:stats}

This appendix details the statistical framework used to infer the
hyperparameters $\Lambda$ describing the BBH population and the
cosmological model. The method follows the hierarchical Bayesian
approach introduced in Ref.~\cite{Mandel:2018mve}.

Single-event parameter estimation is performed in terms of
detector-frame parameters $\theta_D$. The data for each event $i$,
denoted $d_i$, are represented through posterior samples drawn from
$p(\theta_{D,i}|d_i)$ obtained using a reference prior
$\pi_{\rm PE}(\theta_D)$.

For population inference we instead work with source-frame parameters
$\theta$ (component masses, spins, and redshift). These are related to
the detector-frame quantities through a deterministic mapping
$\theta=\theta(\theta_D;\Lambda_c)$ 
which depends on the assumed cosmology via the
distance–redshift relation and the conversion between
detector- and source-frame masses.
Changing variables introduces the Jacobian $J(\theta;\Lambda_c)=\left|\partial\theta_D / \partial\theta \right|$. 

\paragraph{Catalog likelihood}

Let $p_{\rm pop}(\theta|\Lambda)$ denote the population model for the
source parameters. Assuming statistical independence of events, the
hierarchical posterior for a catalog of $N_{\rm obs}$ detections is

\begin{equation}
\begin{split}
p(\{\theta_{i}\}, \Lambda \mid \{d_i\})
\propto \pi(\Lambda) \,
\xi(\Lambda)^{-N_{\rm obs}}
\prod_{i=1}^{N_{\rm obs}}
\frac{p(\theta_{D,i}|d_i)}{\pi_{\rm PE}(\theta_{D,i})}
\\
\times
p_{\rm pop}\!\left(\theta_i(\theta_{D,i};\Lambda_c)\mid\Lambda\right)
J^{-1}\!\left(\theta(\theta_{D,i};\Lambda_c),\Lambda_c\right),
\end{split}
\label{eq:hier_like_basic_full}
\end{equation}

where $\pi(\Lambda)$ denotes the hyperprior. The overall merger-rate
normalization has been analytically marginalized assuming a
scale-invariant prior.
Waveform parameters not explicitly modeled in $p_{\rm pop}$ are
effectively assumed to follow the same prior adopted in the
single-event parameter estimation.

We sample the hierarchical posterior in Eq.~\ref{eq:hier_like_basic_full} directly in the joint
parameter space $\{ \{\theta_i\}, \Lambda \}$~\cite{Mancarella:2025uat}.  
In this approach the event-level posterior
$p(\theta_{D,i}|d_i)$ is approximated with a Gaussian Mixture Model, and the number of mixture components is determined by minimizing the
Bayesian Information Criterion, yielding a smooth representation of
the posterior density suitable for hierarchical sampling.
Further implementation details are provided in
Ref.~\cite{Mancarella:2025uat,Agapito:2026kak}. A blinded mock data challenge
demonstrating the equivalence of this strategy with alternative
likelihood-evaluation procedures is presented in
Ref.~\cite{Tagliazucchi:2026dpr}.  

\paragraph{Likelihood marginalization}

In many applications it is convenient to marginalize
Eq.~(\ref{eq:hier_like_basic_full}) over the event-level parameters.
The resulting likelihood for the hyperparameters alone becomes

\begin{equation}
\begin{split}
p(\{d_i\}\mid \Lambda)
\propto \pi(\Lambda) \,
\xi(\Lambda)^{-N_{\rm obs}}
\prod_{i=1}^{N_{\rm obs}} \!\int \dd \theta_D\, p(d_i|\theta_D)
\\
\times
p_{\rm pop}\!\left(\theta(\theta_D;\Lambda_c)\mid\Lambda\right)
J^{-1}\!\left(\theta(\theta_D;\Lambda_c),\Lambda_c\right).
\end{split}
\label{eq:hier_like_basic_app}
\end{equation}

The integrals over $\theta_D$ are evaluated using posterior samples
from the single-event analyses. If
$\{\theta_{ij}\}_{j=1}^{n_i}$ denotes the sample set for event $i$, the
integral can be approximated through Monte Carlo reweighting as

\begin{equation}
\begin{split}
&\int d\theta_{D, i} \,
p(d_i|\theta_{D, i})
p_{\rm pop}(\theta_i(\theta_{D, i};\Lambda_c)|\Lambda)
J^{-1}(\theta_i(\theta_{D, i};\Lambda_c),\Lambda_c)
\\
&\approx
\frac{1}{n_i}
\sum_{j=1}^{n_i}
\frac{
p_{\rm pop}(\theta_{ij}(\theta_{D,ij};\Lambda_c)|\Lambda)
}{
\pi_{\rm PE}(\theta_{D, ij})\,
J(\theta_{ij}(\theta_{D,ij};\Lambda_c),\Lambda_c)
}
\equiv \hat{\mathcal{L}}_i(\Lambda).
\end{split}
\label{eq:reweight_event_app_integral}
\end{equation}
We adopt this approach as a further validation test.
We retain
9500 posterior samples per event to evaluate these integrals. 

\paragraph{Selection effects}

The factor $\xi(\Lambda)$ represents the detection efficiency, namely
the fraction of sources drawn from the population model that would be
identified by the search pipeline.
We estimate this quantity using simulated injections
\cite{Tiwari:2017ndi,Farr:2019rap}. If injections $\theta_{D,k}$ are
drawn from an injection distribution $\pi_{\rm inj}(\theta_D)$, the
Monte Carlo estimator takes the form
\begin{equation}
\hat{\xi}(\Lambda)=
\frac{1}{N_{\rm inj}}
\sum_{k=1}^{N_{\rm det,inj}}
\frac{
p_{\rm pop}(\theta_k(\theta_{D,k};\Lambda_c)|\Lambda)
}{
\pi_{\rm inj}(\theta_{D,k})\,
J(\theta(\theta_{D,k};\Lambda_c),\Lambda_c)
}.
\label{eq:alpha_inj_app}
\end{equation}
The sum runs over injections recovered by the search pipeline.

\paragraph{Monte Carlo uncertainty}

Both the event-level integrals and the selection-function estimator
are computed via Monte Carlo sums, which introduce stochastic noise
in the likelihood estimate
\cite{Farr:2019rap,Essick:2022ojx,Talbot:2023pex,Heinzel:2025ogf}.
Following Refs.~\cite{Talbot:2023pex,Heinzel:2025ogf}, the variance of
the log-likelihood can be approximated as

\begin{equation}
\sigma^2_{\ln\hat{\mathcal{L}}}(\Lambda)
\simeq
\sum_{i=1}^{N_{\rm obs}}
\frac{\sigma^2_{\hat{\mathcal{L}}_i}}{\hat{\mathcal{L}}_i^2}
+
N_{\rm obs}^2
\frac{\sigma^2_{\hat{\xi}}}{\hat{\xi}^2}.
\end{equation}

Here $\hat{\mathcal{L}}_i(\Lambda)$ denotes the estimator of the
$i$th event likelihood in
Eq.~(\ref{eq:reweight_event_app_integral}), with variance
$\sigma^2_{\hat{\mathcal{L}}_i}(\Lambda)$, while $\hat{\xi}(\Lambda)$
and $\sigma^2_{\hat{\xi}}(\Lambda)$ correspond to the injection-based
selection estimator and its variance. When the full likelihood of
Eq.~\ref{eq:hier_like_basic_full} is sampled directly, only the second
term contributes.

To avoid biases from excessive Monte Carlo fluctuations, we check that
the commonly adopted requirement
\cite{Talbot:2023pex,Heinzel:2025ogf,LIGOScientific:2025pvj} $\sigma^2_{\ln\hat{\mathcal{L}}}(\Lambda)<1$ remains respected.

\section{Prior distributions}
\label{app:priors}

We summarize in Table~\ref{tab:priors} the adopted prior distributions. For compactness, we introduce the notation
$\mathcal{B}_{\sigma}[a,b]$ for the transformed prior
\begin{equation}
X=a+(b-a)\operatorname{sigmoid}(u),
\qquad u\sim\mathcal{N}(0,\sigma^2).
\end{equation}
Similarly, $\mathcal{B}^{\log}_{\sigma}[a,b]$ denotes
\begin{equation}
X=\exp\left[
\ln a+(\ln b-\ln a)\operatorname{sigmoid}(u)
\right],
\qquad u\sim\mathcal{N}(0,\sigma^2).
\end{equation}
These priors have bounded support but are not uniform.

We denote by $\mathcal{N}_{95}[a,b]$ a Gaussian
distribution whose central $95\%$ interval is $[a,b]$.
The notation $\mathcal{L}_{r}[a,b]$ represents a
shifted log-normal distribution with lower bound $a$,
median $a+r(b-a)$, and $95$th percentile $b$.
Finally, $\mathrm{U}[a,b]$ denotes a uniform prior.

\begin{table}[t]
\centering
\footnotesize
\setlength{\tabcolsep}{4pt}
\renewcommand{\arraystretch}{1.08}

\begin{tabular}{@{}llll@{}}
\toprule
Parameter & Prior & Parameter & Prior \\
\midrule

\multicolumn{4}{c}{\emph{Cosmology and GW propagation}} \\
\midrule

$H_0$
& $\mathcal{B}_{1.5}[40,100]$
& $\Omega_m$
& $\mathcal{B}_{1}[0.05,0.55]$
\\
$w_0$
& $\mathcal{B}_{1}[-1.5,-0.5]$
\\

$\Xi_0$
& $\mathcal{B}^{\log}_{1.5}[0.05,5]$
& $n$
& $\mathcal{B}_{1}[0.05,5]$ \\

\midrule
\multicolumn{4}{c}{\emph{Merger-rate distribution}} \\
\midrule

$\gamma$
& $\mathcal{B}_{1.5}[0,6]$
& $\kappa$
& $\mathcal{B}_{1}[0,6]$ \\

$z_p$
& $\mathcal{B}_{1}[0,4]$
& $R_0$
& Marginalized \\

\midrule
\multicolumn{4}{c}{\emph{Primary-mass distribution}} \\
\midrule

$\alpha_{\rm bar}$
& $\mathcal{N}_{95}[1.01,8]$
& $\alpha_{\rm diff}$
& $\mathcal{N}(0,\sqrt{2}\sigma_\alpha)$ \\

$m_b$
& $\mathcal{B}_{1}[20,50]$
& $\mu_1$
& $\mathcal{B}_{1.25}[5,15]$ \\

$\sigma_1$
& $\mathcal{L}_{0.2}[0.1,10]$
& $\mu_2$
& $\mathcal{B}_{1.25}[25,55]$ \\

$\sigma_2$
& $\mathcal{L}_{0.3}[0.1,15]$
& $\delta m_1$
& $\mathcal{L}_{0.3}[0.1,10]$ \\

$m_{1,\min}$
& $3+7u^{3/2}$
& $m_{1,\max}$
& $70+\delta$ \\

$\boldsymbol{\lambda}$
& $\mathrm{Dirichlet}(1,1,1)$
\\

\midrule
\multicolumn{4}{c}{\emph{Secondary-mass distribution}} \\
\midrule

$\beta$
& $\mathcal{N}_{95}[-2,7]$
& $m_{2,\min}$
& $3+v(m_{1,\min}-3)$ \\

$\delta m_2$
& $\mathcal{L}_{0.3}[0.1,10]$
& &
\\

\bottomrule
\end{tabular}

\caption{
Prior distributions for the GP and $\Xi_0$--$n$ models,
using the notation defined in the text.
The GP amplitude and smoothness priors are specified
in Sec.~II B.
The mass and merger-rate entries describe the
reparametrized sampling branches.
}
\label{tab:priors}
\end{table}

For the primary-mass slopes we use
$\alpha_{1,2}=\alpha_{\rm bar}\mp\alpha_{\rm diff}/2$,
where
$\sigma_\alpha=(8-1.01)/(2\times1.95996)$.
The auxiliary variables $u$ and $v$ follow independent
$\mathcal{B}_1[0,1]$ distributions.
For the upper-mass cutoff,
$\delta\sim\mathrm{LogNormal}(\ln65,\ln2/1.95996)$.
This gives a prior median of $135\,M_\odot$ and a
$95$th percentile of $200\,M_\odot$ for $m_{1,\max}$.

\end{document}